\documentclass[sigconf,screen,nonacm,10pt]{vowacmart}

\setcopyright{none}
\renewcommand\footnotetextcopyrightpermission[1]{}

\usepackage{amsmath}
\usepackage{amssymb}
\usepackage{amsthm}
\usepackage{array}
\usepackage{bm}
\usepackage{booktabs}
\usepackage{caption}
\usepackage{enumitem}
\usepackage{graphicx}
\usepackage{makecell}
\usepackage{mathtools}
\usepackage{multirow}
\usepackage{siunitx}
\usepackage{soul}
\usepackage{threeparttable}
\usepackage{tikz}
\usepackage{wasysym}
\usepackage{xurl}

\usetikzlibrary{fit,calc,positioning,arrows.meta}

\microtypecontext{spacing=nonfrench}

\newtheorem{theorem}{Theorem}
\newtheorem{assumption}{Assumption}
\newtheorem{definition}{Definition}

\DeclareFontFamily{U}{stix2bb}{}
\DeclareFontShape{U}{stix2bb}{m}{n} {<-> stix2-mathbb}{}

\NewDocumentCommand{\indicator}{}{\text{\usefont{U}{stix2bb}{m}{n}1}}

\definecolor{lightgreen}{HTML}{d0ffc8}
\definecolor{lightred}{HTML}{fecdc4}
\newcommand{\hlgreen}[1]{{\sethlcolor{lightgreen}\hl{#1}}}
\newcommand{\hlred}[1]{{\sethlcolor{lightred}\hl{#1}}}

\newcolumntype{C}[1]{>{\centering\arraybackslash}m{#1}}

\AtBeginDocument{\clubpenalty=10000
  \widowpenalty=10000
  \displaywidowpenalty=10000
  \finalhyphendemerits=100000
}

\begin{document}

\title{VOW: Verifiable and Oblivious Watermark Detection for Large Language Model Outputs}

\author{Xiaokun Luan}
\authornote{E-mail: \href{mailto:xiaokun.luan@nus.edu.sg}{xiaokun.luan@nus.edu.sg}.}
\affiliation{\institution{National University of Singapore}\city{Singapore}\country{Singapore}}

\author{Yihao Zhang}
\affiliation{\institution{Peking University}\city{Beijing}\country{China}}

\author{Pengcheng Su}
\affiliation{\institution{Peking University}\city{Beijing}\country{China}}

\author{Feiran Lei}
\affiliation{\institution{Peking University}\city{Beijing}\country{China}}

\author{Kailong Wang}
\affiliation{\institution{National University of Singapore}\city{Singapore}\country{Singapore}}

\author{Guangdong Bai}
\affiliation{\institution{City University of Hong Kong}\city{Hong Kong}\country{China}}

\author{Meng Sun}
\authornote{Corresponding author.}
\affiliation{\institution{Peking University}\city{Beijing}\country{China}}

\renewcommand{\shortauthors}{Luan et al.}

\makeatletter
\def\@mkauthors@iii{\gdef\@currentauthors{}\gdef\@currentaffiliation{}\global\setbox\mktitle@bx=\vbox{\unvbox\mktitle@bx\par\medskip
    \centering
    {\@authorfont
      Xiaokun Luan\textsuperscript{1}\footnotemark[1],
      Yihao Zhang\textsuperscript{2},
      Pengcheng Su\textsuperscript{2},
      Feiran Lei\textsuperscript{2}\par
      Kailong Wang\textsuperscript{1},
      Guangdong Bai\textsuperscript{3},
    Meng Sun\textsuperscript{2}\footnotemark[2]\par}
    \smallskip
    {\@affiliationfont
      \textsuperscript{1}National University of Singapore\par
      \textsuperscript{2}Peking University\quad
    \textsuperscript{3}City University of Hong Kong\par}
    \medskip
  }}
\makeatother

\begin{abstract}
  Regulation requires generated text to be marked and detectable, but symmetric-key watermarks leave detection to the key-holding provider.
  We study first-party provenance audit (FPA), where an organization examines sensitive content in its custody.
  Such an audit should not disclose the audited text and should provide a verifiable result and checkable record.
  We present VOW, which uses a verifiable oblivious pseudorandom function to evaluate the detector's keyed component without revealing its inputs to the detection service, while the provider retains the watermarking key.
  A signed setup record published before use fixes the public key and detector specification, and a proof binds returned evaluations to that key.
  The auditor computes the statistic and decision locally, so the service may refuse but cannot make an incorrect result pass verification.
  Given the text, transcript, and setup record, a reviewer can reconstruct the prescribed result without the watermarking key or provider interaction.
  With the detector configured for a nominal false-positive rate of $10^{-5}$, VOW detects 99.4\% of 200-token watermarked texts with competitive text quality and practical overhead.
  Under explicit conditions, our oracle-assisted text construction produces detector-positive text for all three Green-Red detectors evaluated.
  Paraphrasing substantially weakens every scheme evaluated, limiting watermark-based FPA to relatively~intact~content.
\end{abstract}

\maketitle
\setlength{\parfillskip}{0pt plus .75\columnwidth}

\section{Introduction}
\label{sec:introduction}

Large language models (LLMs), such as GPT-4~\cite{gpt4} and Llama~\cite{llama2}, have demonstrated strong capabilities, but their widespread deployment raises questions about the provenance and accountability of generated content.
LLMs can be misused to generate harmful content at scale, ranging from disinformation campaigns~\cite{sun2024exploring} and sophisticated fraud~\cite{lin2024malla} to academic dishonesty and training data contamination~\cite{shumailov2024collapse}.
Opaque LLM APIs also enable model downgrading, in which users unknowingly receive outputs from inferior models~\cite{zhu2025auditing,cai2025are}.
Regulators have responded by making provenance marking a legal requirement.
The EU AI Act requires generative-system providers to mark synthetic text and ensure that it is detectable as artificially generated~\cite{eu_2024_1689_art50}.
California further requires large providers to offer a free, publicly accessible detection tool for image, video, and audio content~\cite{ca_sb942}.
Major providers have meanwhile begun deploying provenance tools such as Google's SynthID~\cite{dathathri2024scalable}.

These requirements place the detection interface on the watermarking system's trust boundary.
A keyed watermark conveys nothing to a party that does not hold the key, so an obligation to make marked content detectable is also an obligation to operate a service that answers detection queries in practice.
Such a service is queried by other parties about content its operator may prefer not to acknowledge, and its answers are consumed as evidence.
We study the security properties of this interface.

Generative watermarking embeds a hidden signal directly into generated content.
While the concept applies to various modalities, its most prominent development has been in text, pioneered by Kirchenbauer et al.~\cite{kirchenbauer2023watermark}.
Their Green-Red scheme (also called the KGW scheme) partitions the vocabulary at each generation step into a favored ``green list'' and a disfavored ``red list'' derived from a secret key.
By biasing the model toward green tokens, the scheme embeds a statistical signal that can be detected later using the same key.

Following this foundational work, subsequent research has largely optimized watermarking for \emph{third-party attribution}, where an independent party holding content it did not produce seeks to assign responsibility for it~\cite{liu2024survey,zhao2025sok}.
Robustness dominates in that setting, because the content may be controlled by an adversary who wishes to escape attribution.
This emphasis has driven developments in semantic watermarking methods designed to survive paraphrasing and rewriting~\cite{hou2024semstamp,hou2024ksemstamp,ren2024robust}.

Deployed practice, however, exposes a different problem.
SynthID has watermarked over ten billion pieces of content~\cite{kohli2025synthiddetector}, but its detector is a Google-operated portal to which the content under examination is uploaded.
For an organization auditing sensitive records, such an interface imposes two costs: it exposes the plaintext to the provider and leaves the auditor unable to verify the returned result independently.

We study this interface in one setting, which we call \emph{first-party provenance audit} (FPA).
The name follows the standard classification of audits by the party conducting them~\cite{iso19011}.
A first-party audit is one an organization conducts on itself, as distinct from one conducted by a customer (second-party) or by an independent body (third-party).
In an FPA, an organization audits selected artifacts already in its custody to determine whether they carry the signal of a marked generative system.
The result informs the organization's own policy and record-keeping decisions, for example when applying AI-assistance disclosure policies to internal reports, proprietary code, or other business records.

An audit of this kind places three requirements on the detection interface.
First, the auditor should not have to disclose the audited text to the provider, since it may contain personal data or confidential information.
Second, because the provider holds the key and computes the result, the auditor should be able to verify it independently.
Third, the audit should leave a record that an outside party with access to the text can later check against the provider's prior commitment.
Together, these requirements protect the auditor's content, spare the provider from receiving plaintext and having to secure it, and support verification both during the audit and in later review.

Current watermarking schemes address these requirements in different ways.
Asymmetric schemes~\cite{christ2024pseudorandom,fairoze2025publicly} use a secret key for watermark insertion and a public key for local detection.
They can satisfy all three interface requirements without involving the provider at detection time, but their cryptographic payloads require long text sequences, often hundreds of tokens, limiting their use for short outputs such as code completion and summarization.
Learning-based schemes~\cite{liu2024unforgeable} also keep detection local by releasing a neural network detector, but do not cryptographically bind it to the marking procedure or provide false-positive control at a specified nominal level.
Proof-based schemes such as PVMark~\cite{duan2026pvmark} make keyed detection publicly verifiable but treat the audited text as public input to the proof.

In this paper, we present VOW, a protocol for \textbf{V}erifiable and \textbf{O}blivious \textbf{W}atermark detection that keeps the auditor's detector inputs private and makes accepted results independently checkable while the provider retains the secret watermarking key.
VOW instantiates the watermarking PRF with a verifiable oblivious pseudorandom function (VOPRF)~\cite{davidson2018privacy,rfc9497}, whose public key and detector specification are fixed in a signed, versioned setup record published before use.
The auditor blinds each unique context-token pair and obtains its keyed evaluation without disclosing the pair to the detection service.
The service returns a batched proof binding its evaluations to the committed key, and the auditor derives the pair colors and computes the statistic and decision locally.
The service may refuse, but it cannot make an incorrect result pass verification.
A later reviewer given the text, transcript, and setup record can reconstruct the prescribed detector result without the watermarking key or further provider interaction.
A positive result supports provenance only if use of the watermarking key is confined to the declared pipeline and the artifact producer did not assemble the text using detector feedback.
We also examine oracle-assisted text construction and the loss of detectability under rewriting.
At the nominal significance level $\alpha=10^{-5}$, VOW produces six false positives among one million unwatermarked texts and reaches a TPR of 0.994 on 200-token watermarked texts, with competitive text quality and practical insertion and detection overhead.

In summary, our main contributions are threefold:
\begin{enumerate}[noitemsep, leftmargin=*]
  \item We formalize FPA against a malicious detection service and define input privacy, detection soundness, and transferable evidence relative to a previously published setup record.
  \item We design and implement VOW using a VOPRF and a statistical test run by the auditor, establish that it realizes these goals, and evaluate its detectability, text quality, and overhead on short texts.
  \item We characterize two boundaries of watermark-based FPA: oracle-assisted text construction under explicit interface conditions, and loss of detectability under rewriting.
    The construction produces detector-positive text for all three Green-Red detectors evaluated, while paraphrasing substantially weakens every scheme evaluated.
\end{enumerate}

\section{Preliminaries}
\label{sec:preliminaries}

This section introduces notation for language models and the core cryptographic primitive that underpins our work.

\subsection{Language Model Notation}

We consider an autoregressive language model over a vocabulary $\mathcal{V}=\{w_1,\ldots,w_N\}$ of size $N$.
Given a sequence of tokens $(t_1,\ldots,t_n)\in\mathcal{V}^{n}$, the model outputs logit scores $l\in\mathbb{R}^N$ for the next token $t_{n+1}$, where $l_i$ corresponds to the $i$-th token $w_i \in \mathcal{V}$.
A softmax transforms these logits into a probability distribution $p\in\Delta^{N}$, from which the next token $t_{n+1}$ is sampled.
While standard generation samples directly from $p$, practical deployments often use top-k or top-p (nucleus) truncation to improve quality by restricting sampling to a subset of $\mathcal{V}$.

\subsection{Verifiable Oblivious PRFs}

\begin{figure*}[!t]
    \centering

    \definecolor{accent}{HTML}{1F4E79}
    \definecolor{softgray}{HTML}{6E6E6E}
    \definecolor{lightgray}{HTML}{FAFAFA}
    \definecolor{notegray}{HTML}{737373}
    \definecolor{alertred}{HTML}{8A3B2F}
    \definecolor{stepbg}{HTML}{F4F7FB}

    \tikzset{
        pics/keyglyph/.style n args={1}{code={
                \draw[line width=0.4pt, fill=#1] (0,0) circle (0.078);
                \draw[line width=0.55pt] (0.078,0) -- (0.34,0);
                \draw[line width=0.55pt] (0.22,0) -- (0.22,-0.075);
                \draw[line width=0.55pt] (0.32,0) -- (0.32,-0.075);
        }},
        pics/person/.style={code={
                \draw[line width=0.45pt] (0,0.11) circle (0.085);
                \draw[line width=0.45pt] (-0.145,-0.15) arc (180:0:0.145 and 0.165);
        }},
        pics/server/.style={code={
                \draw[line width=0.45pt] (-0.15,-0.01) rectangle (0.15,0.12);
                \draw[line width=0.45pt] (-0.15,-0.19) rectangle (0.15,-0.06);
                \fill (-0.10,0.055) circle (0.018);
                \fill (-0.10,-0.125) circle (0.018);
        }},
        pics/scales/.style={code={
                \draw[line width=0.45pt] (0,-0.15) -- (0,0.13);
                \draw[line width=0.45pt] (-0.16,0.11) -- (0.16,0.11);
                \draw[line width=0.45pt] (-0.16,0.11) -- (-0.23,0.00) -- (-0.09,0.00) -- cycle;
                \draw[line width=0.45pt] (0.16,0.11) -- (0.09,0.00) -- (0.23,0.00) -- cycle;
                \draw[line width=0.45pt] (-0.09,-0.15) -- (0.09,-0.15);
        }},
        party/.style={draw, rounded corners=2.5pt, align=center,
        font=\small, inner sep=4pt},
        optional/.style={party, dashed, dash pattern=on 2pt off 1.6pt},
        orgbox/.style={draw=softgray, dotted, thick, rounded corners=6pt},
        msg/.style={-{Stealth[length=4.5pt,width=3.5pt]}, semithick, accent},
        content/.style={-{Stealth[length=4pt,width=3pt]}, line width=0.55pt, softgray,
        dash pattern=on 2.6pt off 1.8pt},
        note/.style={font=\footnotesize, align=center, inner sep=1.5pt},
        sub/.style={font=\scriptsize\itshape, color=softgray, align=center, inner sep=1.5pt},
    }

    \providecommand{\stepdisc}[2]{\tikz[baseline=#2]{\node[circle, fill=accent, text=white,
    inner sep=0pt, minimum size=2.5mm, font=\tiny\bfseries] {#1};}}
    \providecommand{\subline}[1]{{\scriptsize\itshape\color{notegray} #1}}
    \providecommand{\keyfilled}{\tikz[baseline=-0.7ex]{\pic at (0,0) {keyglyph=black};}}
    \providecommand{\keyhollow}{\tikz[baseline=-0.7ex]{\pic at (0,0) {keyglyph=white};}}
    \providecommand{\arrmsg}{\tikz[baseline=-0.4ex]{\draw[msg] (0,0) -- (0.36,0);}}
    \providecommand{\arrcnt}{\tikz[baseline=-0.4ex]{\draw[content] (0,0) -- (0.36,0);}}

    \begin{tikzpicture}[x=1cm, y=1cm]

        \node[party, minimum width=4.5cm, minimum height=0.8cm] (O) at (1.8,3.15)
        {Model Owner $\mathcal{O}$};
        \node[party, minimum width=4.5cm, minimum height=0.9000cm] (D) at (1.8,1.40)
        {Detection Service $\mathcal{D}$\\[-2pt]{\scriptsize\itshape\color{alertred}may be malicious}};

        \node[orgbox, fit=(O)(D), inner sep=5pt] (ORG) {};
        \node[sub, anchor=south] at (ORG.north) {provider organization};

        \node[party, minimum width=3.0cm, minimum height=0.9000cm] (A) at (10.2,1.40)
        {Auditor $\mathcal{A}$\\[-2pt]{\scriptsize\itshape\color{softgray}holds candidate $T$}};
        \node[optional, minimum width=2.8cm, minimum height=0.9000cm] (R) at (15.3,1.40)
        {Arbiter\\[-2pt]{\scriptsize\itshape\color{softgray}optional}};

        \pic at ($(O.north west)+(0.30,-0.24)$) {server};
        \pic at ($(D.north west)+(0.30,-0.24)$) {server};
        \pic at ($(A.north west)+(0.30,-0.24)$) {person};
        \pic at ($(R.north west)+(0.30,-0.24)$) {scales};

        \pic at ($(O.north east)+(-0.56,-0.22)$) {keyglyph=black};
        \pic at ($(D.north east)+(-0.56,-0.22)$) {keyglyph=black};
        \pic at ($(A.north east)+(-0.56,-0.22)$) {keyglyph=white};
        \pic at ($(R.north east)+(-0.56,-0.22)$) {keyglyph=white};

        \draw[msg] ($(O.south)+(-11mm,0)$) -- node[note, right, xshift=1pt] {\stepdisc{1}{-0.6ex}~shares $\mathsf{wk}=\mathsf{sk}$\\[1pt]
        \subline{signed $\mathsf{SR}$ published; contains $\mathsf{pk}$}} ($(D.north)+(-11mm,0)$);

        \draw[content] (O.east) -- (7.0,3.15)
        node[note, above, color=softgray, pos=1.12] {marked output $T$}
        node[sub, below, pos=1.12] {ordinary use --- the path an audit tests}
        -| (A.north);

        \draw[msg] ($(A.west)+(0,0.28)$) --
        node[note, above, color=accent] {\stepdisc{2}{0.6ex}~$n$ blinded inputs $\tilde{x}_i$\\[1pt]
        \subline{cryptographic payload reveals only $n$}}
        ($(D.east)+(0,0.28)$);
        \draw[msg] ($(D.east)+(0,-0.29)$) --
        node[note, above, color=accent, align=center] {\stepdisc{3}{0.6ex}~blinded values $\tilde{y}_i$ with proof $\pi$}
        node[sub, below] {bound to the committed key $\mathsf{pk}$}
        ($(A.west)+(0,-0.29)$);

        \draw[msg] (A.east) --
        node[note, above, color=accent] {\stepdisc{5}{0.6ex}~$T,\tau,\mathsf{SR}$}
        node[sub, below] {re-checkable}
        (R.west);

        \node[draw=accent!55, line width=0.5pt, fill=stepbg, rounded corners=2pt, align=left, font=\footnotesize, inner sep=4pt, anchor=west, align=center] (RES) at (12.1,2.85)
        {\color{accent}\stepdisc{4}{-0.6ex}~verifies $\pi$ under $\mathsf{pk}$, else abort\\[1pt]
        \subline{run the statistical test locally to get $p$-value}};
        \draw[softgray, line width=0.4pt] ($(A.north east)+(-0.15,0)$) to[out=70,in=200] ($(RES.west)+(-0.02,-0.10)$);

        \node[anchor=north west, font=\scriptsize\itshape, color=notegray, inner sep=1pt, fill=lightgray, fill opacity=0.7, text opacity=1, rounded corners=2pt]
        at ($(RES.north west)+(-4.2,0.76)$)
        {\keyfilled\hspace{1.5mm}holds $\mathsf{wk}$\hspace{4mm}
            \keyhollow\hspace{1.5mm}holds $\mathsf{SR}$\hspace{4mm}
            \arrmsg\hspace{1.5mm}protocol message\hspace{4mm}
            \arrcnt\hspace{1.5mm}content path
        };

    \end{tikzpicture}

    \caption{Trust model and detection flow. The model owner publishes the signed setup record $\mathsf{SR}$ and shares the watermarking key $\mathsf{wk}$ with the detection service; the protocol constrains the service's behavior during an accepted detection execution.}

    \label{fig:overview}
\end{figure*}

An oblivious pseudorandom function (OPRF) is a two-party protocol between a server holding a pseudorandom function (PRF) key $k$ and a client holding an input $x$.
The protocol allows the client to learn the PRF output $\mathcal{F}(k,x)$ without revealing $x$ to the server, while the server learns nothing about the output~\cite{casacuberta2022sok}.
A verifiable OPRF (VOPRF) further lets the client verify that the server evaluated the PRF with a specific, committed key, typically via a proof of correct evaluation~\cite{davidson2018privacy}.
Owing to these two properties, VOPRFs underpin a range of privacy-preserving systems, \mbox{including Privacy Pass~\cite{davidson2018privacy}}.

\begin{definition}
  \label{def:voprf}
  A VOPRF is a tuple of algorithms with security parameter $\lambda$ and output length $\ell=\ell(\lambda)$:
  \begin{itemize}[noitemsep, leftmargin=*]
    \item $\mathsf{Setup}(1^\lambda)\to (\mathsf{sk},\mathsf{pk})$: A key generation algorithm run by the server to get a secret key $\mathsf{sk}$ and a public key $\mathsf{pk}$.
    \item $\mathsf{Evaluate}(\mathsf{sk},x)\to y$: A deterministic server-side algorithm to compute the PRF output $y=\mathcal{F}(\mathsf{sk},x)\in\{0,1\}^\ell$ for an input $x$.
    \item The interactive oblivious evaluation protocol consists of:
      \begin{itemize}[noitemsep, leftmargin=*]
        \item $\mathsf{Blind}(x)\to (\tilde{x}, r)$: A client-side algorithm to produce a blinded input $\tilde{x}$ and a blinding factor $r$.
        \item $\mathsf{BlindEvaluate}(\mathsf{sk}, \tilde{x})\to (\tilde{y},\pi)$: Server evaluation of a blinded input, returning its output and a proof $\pi$ of correct evaluation.
        \item $\mathsf{Verify}(\mathsf{pk},\tilde{x},\tilde{y},\pi)\to \{\mathit{Accept},\mathit{Reject}\}$: A client-side verification algorithm that checks whether $\tilde{y}$ is a valid evaluation of $\tilde{x}$ under the secret key corresponding to $\mathsf{pk}$.
        \item $\mathsf{Finalize}(x, \tilde{y}, r)\to y$: Client recovery of $y\in\{0,1\}^\ell$ \mbox{after verification}.
      \end{itemize}
  \end{itemize}
\end{definition}

We instantiate the VOPRF with the 2HashDH construction~\cite{jarecki2014roundoptimal} from RFC 9497~\cite{rfc9497}.
Our implementation uses its batched proof mode, in which the server evaluates multiple blinded inputs in one invocation and returns the corresponding outputs with a single proof covering the entire batch.
For this mode, we write $\mathsf{VerifyBatch}(\mathsf{pk},\allowbreak\tilde{\mathbf{x}},\allowbreak\tilde{\mathbf{y}},\allowbreak\pi)\to\{\mathit{Accept},\mathit{Reject}\}$ for verification of the batch proof.

\section{Threat Model and Security Goals}
\label{sec:threat-model}

\subsection{Setting and Parties}
\label{sec:tm-setting}

We consider a first-party provenance audit in which an organization examines a text $T$ already in its custody for the signal associated with a provider's declared marked generation pipeline.
The organization uses the result for its own policy and record-keeping decisions.

Figure~\ref{fig:overview} summarizes the audit setting and the roles described below.
The \textbf{Model Owner} $\mathcal{O}$ operates the marked generation pipeline with a secret watermarking key and declares the watermark setup.
The \textbf{Detection Service} $\mathcal{D}$ uses the same key to answer detection queries.
The \textbf{Auditor} $\mathcal{A}$ holds a text $T$ and seeks its detection result without disclosing the text to $\mathcal{D}$.
$\mathcal{O}$ and $\mathcal{D}$ may be operated by the same provider, and we assume no independence between them.
We model $\mathcal{D}$ as malicious because it performs the online computation under the watermarking key and may have an interest in the audit outcome.
It may deviate arbitrarily or refuse service.
The auditor is honest and follows the protocol.
For $\mathcal{O}$, the security goals are conditioned on the valid setup record specified below.

\begin{assumption}[Published setup record]
  \label{asm:setup}
  Before a setup is used for watermark insertion or detection, $\mathcal{O}$ publishes a versioned setup record $\mathsf{SR}$ signed under an authenticated organizational key.
  The record identifies $\mathcal{O}$, commits to the watermarking key, and fixes the detector specification against which subsequent results are computed and verified.
  It also identifies the generation pipeline and outputs that $\mathcal{O}$ declares~it~will~mark.
\end{assumption}

For a valid $\mathsf{SR}$ and a presented text $T$, the detector specification determines the inputs derived from $T$ that are evaluated under the watermarking key to which $\mathsf{SR}$ commits.
Let $\mathsf{Result}_{\mathsf{SR}}(T)$ denote the complete prescribed result, consisting of those inputs, their corresponding outputs, and the resulting statistic and decision.

\subsection{Security Goals}
\label{sec:tm-properties}

Adversaries are probabilistic polynomial-time in the security parameter $\lambda$.
We require input privacy during detection, correctness of the result whenever the auditor completes the protocol, and a record that can be checked later.

\begin{definition}[Input privacy]
  \label{def:privacy}
  For every malicious detection service $\mathcal{D}^{*}$, valid setup record $\mathsf{SR}$, and pair of texts $T_0,T_1$ for which the specified detector performs the same number of keyed evaluations, the cryptographic views of $\mathcal{D}^{*}$ in two protocol executions with an honest auditor holding $T_0$ and $T_1$, respectively, are computationally indistinguishable.
\end{definition}

The cryptographic exchange therefore reveals the number of keyed evaluations but computationally hides further information about $T$.

\begin{definition}[Detection soundness]
  \label{def:soundness}
  For every valid setup record $\mathsf{SR}$, text $T$, and malicious detection service $\mathcal{D}^{*}$, the probability that an honest auditor completes the protocol without aborting and outputs a result different from $\mathsf{Result}_{\mathsf{SR}}(T)$ is negligible in $\lambda$.
\end{definition}

A malicious service may prevent an audit from completing, but it cannot make an incorrect result pass the auditor's checks.
On completion, the auditor retains a transcript $\tau$ that it may later present together with $T$ and $\mathsf{SR}$ to a reviewer.

\begin{definition}[Transferable evidence]
  \label{def:evidence}
  Given $T$, $\tau$, and $\mathsf{SR}$, any reviewer can run an efficient public procedure
  $\mathsf{Arbitrate}(T,\tau,\mathsf{SR})$ without the secret watermarking key or interaction with the provider.
  The procedure either rejects or returns a detection result, and returns one only after authenticating $\mathsf{SR}$.
  On the transcript of any honestly completed audit, it returns $\mathsf{Result}_{\mathsf{SR}}(T)$.
  For any probabilistic polynomial-time adversary that produces $\tau$, the probability that $\mathsf{Arbitrate}$ returns a different result from $\mathsf{Result}_{\mathsf{SR}}(T)$ is negligible in $\lambda$.
\end{definition}

Arbitration is selective disclosure: it requires the presented text, but the auditor chooses whether and to whom that text is disclosed.
Together, detection soundness and transferable evidence ensure that the auditor's accepted result and any result accepted in later review equal $\mathsf{Result}_{\mathsf{SR}}(T)$.

\subsection{Scope Boundaries}
\label{sec:tm-scope}

We next clarify three important boundaries of these security goals that affect how detection results should be interpreted.

\noindent\textbf{Generation conduct and compliance.}
The setup record identifies a generation pipeline and the outputs that the provider declares it will mark.
A positive result is evidence that $T$ came from the declared pipeline only when the watermark key is confined to the declared pipeline and its prescribed detection service.
We call adherence to this declared marking policy \emph{faithful marking}.
Assessing it typically requires sampling outputs covered by the declaration and comparing their detection results with the policy.

\noindent\textbf{Oracle-assisted text construction.}
Without knowing the watermarking key, an artifact producer may use exact feedback from the detection interface to assemble a text that passes an honest audit.
Such a positive result does not distinguish generation by the declared pipeline from text constructed using detector feedback but without the key.
We analyze and evaluate one such construction for Green-Red watermark detection in Section~\ref{sec:analysis-forgery}.

\noindent\textbf{Modification.}
Editing may weaken or erase the watermark signal, so a modified version of a watermarked text may receive a negative result.
Because there is no agreed formal notion of meaning-preserving modification, we treat robustness as an empirical property and evaluate it in Section~\ref{sec:evaluation-robustness}.

\section{The VOW Protocol}
\label{sec:method}

This section presents VOW, a protocol for verifiable and oblivious watermark detection built on the Green-Red watermarking scheme.
VOW replaces the color-determination logic so that it can be evaluated privately and verifiably using a VOPRF.

\subsection{The Green-Red Watermarking Scheme}

In the Green-Red scheme, introduced by Kirchenbauer et al.~\cite{kirchenbauer2023watermark}, the vocabulary $\mathcal{V}$ is pseudorandomly partitioned into a green list $G$ and a red list $R$ at each generation step.
The partition for token $t_k$ is derived from the preceding context $c=(t_{k-h},\ldots,t_{k-1})$ with $h$ tokens and a PRF keyed by a secret watermarking key $\mathsf{wk}$.
In KGW, the PRF output seeds a pseudorandom permutation of the vocabulary, and the first $\gamma N$ entries form the green list, where $\gamma\in(0,1)$ is the nominal green fraction.
Logits of green tokens are raised by a bias $\delta > 0$:
\begin{equation}
  \tilde{l}_{i} =
  \begin{cases}
    l_{i} + \delta, & \text{if } w_i \in G, \\
    l_{i},          & \text{otherwise.}
  \end{cases}
\end{equation}
Detection recomputes the color of each token and tests whether green tokens are overrepresented relative to the nominal rate $\gamma$.
The same key $\mathsf{wk}$ determines color during both insertion and detection.

\subsection{Our Watermarking Protocol: VOW}
\label{sec:method-protocol}

VOW retains the Green-Red structure but derives the color of each context-token pair directly from a keyed PRF, allowing color determination to be evaluated obliviously.
Let $\mathcal{F}:\mathcal{K}\times\mathcal{X}\to\{0,1\}^\ell$ be a PRF keyed by $\mathsf{wk}$.
Interpreting its output as a number in the unit interval, we define the color predicate as:
\begin{equation}
  \mathrm{IsGreen}_\mathsf{wk}(c, t) \Leftrightarrow \frac{\mathrm{Int}(\mathcal{F}(\mathsf{wk},c \Vert t))}{2^\ell} < \gamma,
\end{equation}
where $\mathrm{Int}(\cdot)$ reads the $\ell$-bit output as an unsigned integer, $\Vert$ denotes the canonical encoding of the context-token pair specified by $\mathsf{SR}$, and $\gamma\in (0,1)$ sets the threshold and expected green-token ratio.
The green list $G$ is then the set of tokens $t$ for which $\mathrm{IsGreen}_\mathsf{wk}(c,t)$ holds.

We implement $\mathcal{F}$ using the VOPRF of Section~\ref{sec:preliminaries}, with direct evaluation during insertion and oblivious evaluation during detection.
Obliviousness hides each queried context-token pair from the detection service, and verifiability lets the auditor reject incorrect evaluations.
Figure~\ref{fig:overview} summarizes VOW's deployment and detection flow.
We next describe the three phases of the protocol.

\noindent\textbf{Setup.}
The Model Owner $\mathcal{O}$ runs $\mathsf{Setup}(1^\lambda)\to(\mathsf{sk},\mathsf{pk})$, keeps $\mathsf{sk}$ as the watermarking key $\mathsf{wk}$, makes the same key available to $\mathcal{D}$ for detection, and signs and publishes the setup record $\mathsf{SR}$ of Assumption~\ref{asm:setup}.
Concretely, the detector specification in $\mathsf{SR}$ fixes the protocol version, the VOPRF public key $\mathsf{pk}$ corresponding to $\mathsf{wk}$, the tokenizer and pair encoding, $h$ and $\gamma$, the deduplication rule, and the exact statistical test with significance level $\alpha$.
The record also identifies the generation pipeline and the outputs that the provider declares it will mark.

\noindent\textbf{Insertion.}
Watermark insertion is local to $\mathcal{O}$ and requires neither blinding nor proofs.
For a context $c=(t_1,\ldots,t_h)$, $\mathcal{O}$ evaluates $y_i=\mathsf{Evaluate}(\mathsf{wk},c\Vert w_i)$ for candidate tokens $w_i\in\mathcal{V}$ and shifts each logit by
\begin{equation}
  \tilde{l}_i = l_i + \delta \cdot \indicator\!\left(\frac{\mathrm{Int}(y_i)}{2^\ell} < \gamma\right),
\end{equation}
where $\indicator$ is the indicator function.
The next token $t_{h+1}$ is sampled from the modified logits $\tilde{l}$.
We discuss optimizations to avoid iterating over the entire vocabulary during the sampling process in Section~\ref{sec:method-optimizations}.

\noindent\textbf{Detection.}
Before issuing detection queries, $\mathcal{A}$ verifies $\mathcal{O}$'s signature on $\mathsf{SR}$ and checks that its protocol version, public key, and detector specification are supported and well formed.
The detection phase is an interactive protocol between the Auditor $\mathcal{A}$, acting as client, and the Detection Service $\mathcal{D}$, acting as server.
Using the detector specification in $\mathsf{SR}$, $\mathcal{A}$ first derives the unique context-token pairs $x_1,\ldots,x_n$, where $n$ is their number and $x_i=c_i\Vert t_i$.
The protocol then proceeds in the following steps:
\begin{enumerate}[noitemsep, leftmargin=*]
  \item \emph{Blinding}: $\mathcal{A}$ runs $(\tilde{x}_i,r_i)=\mathsf{Blind}(x_i)$ to obtain blinded inputs and sends the vector $\tilde{\mathbf{x}}=(\tilde{x}_1,\ldots,\tilde{x}_n)$ to $\mathcal{D}$.
  \item \emph{Evaluation}: $\mathcal{D}$ returns $(\tilde{\mathbf{y}},\pi)=\mathsf{BlindEvaluate}(\mathsf{wk},\tilde{\mathbf{x}})$.
  \item \emph{Verification}: $\mathcal{A}$ checks $\mathsf{VerifyBatch}(\mathsf{pk},\tilde{\mathbf{x}},\tilde{\mathbf{y}},\pi)$ using the public key in $\mathsf{SR}$ and aborts if it rejects.
  \item \emph{Finalization}: $\mathcal{A}$ computes $y_i=\mathsf{Finalize}(x_i,\tilde{y}_i,r_i)$ for every $i$ and classifies $(c_i,t_i)$ as green if $\mathrm{Int}(y_i)/2^\ell < \gamma$.
\end{enumerate}

Following Fernandez et al.~\cite{fernandez2023three}, $\mathcal{A}$ counts each unique pair once because repeated pairs share the same color and are not independent trials.
Let $g$ denote the observed green pair count and $X$ the corresponding random variable.
For an unwatermarked $T$ selected independently of the committed key and detection oracle, the null model is $X\sim\mathrm{Binomial}(n,\gamma)$, against the alternative that its green proportion exceeds $\gamma$.
A one-sided exact binomial test gives
\begin{equation}
  p(g;n,\gamma)
  =\Pr_{X\sim\mathrm{Binomial}(n,\gamma)}[X\ge g]
  =I_\gamma(g,n-g+1),
\end{equation}
where $I_x(a,b)$ is the regularized incomplete beta function.
The decision is positive exactly when this $p$-value is at most the significance level $\alpha$ specified in $\mathsf{SR}$.

Finally, the auditor records $g$ and the resulting decision $d$, together with the VOPRF exchange, blinding data, and an identifier for the selected $\mathsf{SR}$, as the transcript $\tau$.

\subsection{Design Rationale}
\label{sec:method-rationale}

We discuss two design choices that make oblivious detection feasible at vocabulary scale.

The first is the pointwise color determination logic, which departs from the permutation-based green list of KGW.
Evaluating the permutation-based rule obliviously requires oblivious set membership testing, for which private set intersection~\cite{kissner2005privacy} is the natural tool, and its cost grows with the vocabulary size.
Framing color as a direct PRF evaluation on the input $c\Vert t$ removes the membership test entirely and thus brings the predicate within the reach of a VOPRF.

The second is to evaluate this PRF with a VOPRF.
During insertion, $\mathcal{O}$ evaluates it directly, whereas during detection, $\mathcal{D}$ evaluates blinded inputs and proves consistency with the public key in $\mathsf{SR}$.
This lets $\mathcal{A}$ derive colors from verified outputs and run the statistical test locally, while keeping the context-token pairs hidden from $\mathcal{D}$ and $\mathsf{wk}$ hidden from $\mathcal{A}$.

A tradeoff concerns the context length $h$.
Larger contexts make reusable context-token lookup tables more costly but cause a local token change to affect more subsequent pairs.
We use $h=2$ as the default middle ground.

The pointwise color rule makes VOPRF-based detection practical but raises insertion overhead, since each candidate token may require an independent PRF evaluation.
The following optimizations address this overhead.

\subsection{Practical Optimizations}
\label{sec:method-optimizations}

\subsubsection{Efficient and Unbiased Sampling}
Evaluating $\mathrm{IsGreen}$ over the whole vocabulary at each step is computationally prohibitive, especially for large vocabularies.
Precomputation and caching do not help, because the green list $G$ depends on the context.
KGW's SelfHash variant meets the same obstacle and evaluates only a fixed set of high-logit tokens, for instance the top 40 in~\cite{kirchenbauer2023watermark}, which is an approximation that biases the sampled distribution.
We instead use two techniques that reduce cost while preserving the target distribution exactly: rejection sampling for multinomial sampling, and an adaptive sieve with early exit for top-$k$ filtering.

\noindent\textbf{Unbiased multinomial sampling.}
Let $p(t)$ be the model's original probability for token $t$ and $q(t)$ the target probability distribution after applying the bias $\delta$, so that
\begin{equation}
  q(t) \propto
  \begin{cases}
    p(t) \cdot \exp(\delta), & \text{if } t \in G, \\
    p(t),                    & \text{otherwise.}
  \end{cases}
\end{equation}
We sample from $q$ by rejection sampling with $p$ as the proposal: we draw $t\sim p$, accept it if $t\in G$, and otherwise accept it with probability $\exp(-\delta)$, repeating until acceptance.
This rejection rule samples exactly from $q$ (proof provided in Appendix~\ref{app:unbiased-sampling}).
Writing $\rho=\sum_{t\in G}p(t)$, the expected number of trials is $\frac{\exp(\delta)}{1 + (\exp(\delta) - 1) \rho}$.
For $\delta=3.0$ and $\rho\approx 0.5$, as expected when $\gamma=0.5$, this yields about 1.91 trials per token, nearly five orders of magnitude fewer evaluations than a full pass over a vocabulary of roughly \num{152000} entries.

\begin{figure}[t]
  \centering
  \includegraphics{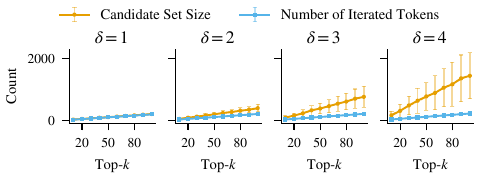}
  \caption{Average candidate set size and number of tokens evaluated before early stopping for different values of the watermark parameter $\delta$ and sampling parameter $k$. Error bars indicate standard deviations over 1000 generation runs.}
  \label{fig:candidate-set-size}
\end{figure}

\noindent\textbf{Accelerated top-$k$ filtering.}
Rejection sampling does not extend to top-$k$ sampling, a common technique for improving generation quality, so we introduce a two-stage filter.

First, we fix a candidate set $C_k$ by the adaptive cutoff logit $u_k = l_{k\text{-th}} - \delta$, where $l_{k\text{-th}}$ is the logit of the $k$-th ranked token before biasing.
Any token below this cutoff cannot enter the new top-$k$ set even with the full bias applied, so the filter is exact rather than heuristic.
Figure~\ref{fig:candidate-set-size} shows that the size of $C_k$ grows linearly in $k$, with a coefficient typically between 2 and 15 depending on $\delta$.

Second, we add an early exit.
We iterate through the candidates in $C_k$ in descending order of their original logits, maintaining a min-heap of the highest $k$ biased scores seen so far.
After processing $w_j$, we compare the heap's minimum $\tilde{l}_{\min}$ against the next candidate's original logit $l_{j'}$.
If $\tilde{l}_{\min} > l_{j'} + \delta$, no subsequent candidate can possibly enter the top-$k$ set and the iteration stops.
Figure~\ref{fig:candidate-set-size} shows that this reduces the average number of $\mathrm{IsGreen}$ calls to roughly $2k$, offering a significant speedup over evaluating the full candidate set.

\subsubsection{Efficient Detection}

Returning one proof per input would make proof communication linear in the number of queried context-token pairs.
We instead use the batching mode of RFC 9497~\cite{rfc9497}, in which $\mathcal{D}$ returns a single proof covering a set of blinded inputs and $\mathcal{A}$ accepts only if batch verification succeeds for all of them~\cite{davidson2018privacy}, reducing proof communication to one proof per batch.
A deployment that requires a separately verified proof for every input can disable batching at the cost of linear proof communication.
The efficiency measures in Section~\ref{sec:evaluation-overhead} use the batched mode.

\section{Security Analysis}
\label{sec:analysis}

This section establishes the three properties defined in Section~\ref{sec:threat-model} and then analyzes oracle-assisted text construction and text modification.

\subsection{Input Privacy}
\label{sec:analysis-privacy}

Input privacy follows from the obliviousness of the VOPRF.
In the 2HashDH instantiation, each input $x_i$ is mapped by the hash-to-group function $H_1$ and blinded as $\tilde{x}_i = H_1(x_i)^{r_i}$.
With uniformly sampled nonzero blinding scalars $r_i$, $\tilde{x}_i$ is independent of $x_i$ whenever the hash-to-group output is not the identity; the standard protocol rejects the negligible identity event.
Consequently, for any two texts inducing the same number $n$ of keyed evaluations, the detection service's protocol views are statistically, and hence computationally, indistinguishable as required by Definition~\ref{def:privacy}.
Theorem~\ref{thm:privacy} in Appendix~\ref{app:security-proof} formalizes this argument.
The cryptographic exchange reveals $n$.
Transport metadata such as account identity, timing, and request boundaries lies outside this view.

\subsection{Detection Soundness}
\label{sec:analysis-soundness}

Detection soundness follows from the soundness of the batched DLEQ proof system and the auditor's deterministic local computation.
Theorem~\ref{thm:soundness} in Appendix~\ref{app:security-proof} gives the formal argument.
Our implementation returns one non-interactive batched discrete-log equality proof~\cite{chaum1992wallet} for the vectors $\tilde{\mathbf{x}}$ and $\tilde{\mathbf{y}}$.
The verifier checks the batch through the constant-size composite proof specified by the batched VOPRF construction.
Under the batched proof system's soundness, a batch containing any blinded input-output pair inconsistent with the public key fixed by $\mathsf{SR}$ is accepted only with negligible probability.
Because the auditor derives the prescribed inputs and computes the statistic and decision locally, every accepted result equals $\mathsf{Result}_{\mathsf{SR}}(T)$ except with negligible probability.

\subsection{Transferable Evidence}
\label{sec:analysis-evidence}

The transcript retained after detection consists of the VOPRF exchange $(\tilde{\mathbf{x}},\tilde{\mathbf{y}},\pi)$, the blinding factors $r_1,\ldots,r_n$, the count and decision $(g,d)$, and an identifier for locating the setup record.
Given $T$, $\tau$, and $\mathsf{SR}$, a reviewer performs the following checks.
\begin{enumerate}[noitemsep, leftmargin=*]
  \item Authenticate $\mathsf{SR}$, confirm that it is the record identified in $\tau$, and extract its public key and detector specification.
  \item Derive the unique inputs $x_1,\ldots,x_n$ from $T$ under that specification.
    Require a one-to-one match between these inputs and the recorded triples $(r_i,\tilde{x}_i,\tilde{y}_i)$, confirm that every $r_i$ is valid blinding randomness, and recompute the blinded input from $x_i$ and $r_i$ to confirm that it equals $\tilde{x}_i$.
  \item Verify $\pi$ for $(\tilde{\mathbf{x}},\tilde{\mathbf{y}})$ under the public key in $\mathsf{SR}$.
  \item Compute $y_i=\mathsf{Finalize}(x_i,\tilde{y}_i,r_i)$ for every $i$, recompute the green count, $p$-value, and decision under the detector specification, and confirm that the recorded $(g,d)$ agrees.
\end{enumerate}

On an honest transcript, these checks accept and reconstruct $\mathsf{Result}_{\mathsf{SR}}(T)$.
For any accepted transcript, the second check ties every input derived from $T$ to exactly one proof-covered entry.
Except with negligible probability, batched-proof soundness binds each corresponding output to the public key in $\mathsf{SR}$.
Finalization and the last check then reconstruct the complete prescribed result.
All checks use only $T$, $\tau$, $\mathsf{SR}$, and public algorithms.
They require neither the watermarking key nor interaction with the provider.
Theorem~\ref{thm:evidence} in Appendix~\ref{app:security-proof} formalizes this argument.

\subsection{Oracle-Assisted Text Construction}
\label{sec:analysis-forgery}

VOW lets the auditor recover the color of each queried context-token pair and run the statistical test locally.
We show how an artifact producer without $\mathsf{wk}$ can use the feedback from the detection interface under a fixed $\mathsf{SR}$ to select tokens adaptively and construct text outside the declared generation pipeline.
If successful, the construction produces a text that receives a positive decision in an honest audit and can yield a valid transcript.
The setup signature and VOPRF proofs verify, and the audit returns the prescribed detector result.
The positive result nevertheless cannot distinguish text produced by the declared pipeline from text assembled outside it.

The construction fixes a candidate budget $k\ge 1$ at every position and begins with an initial context of $h$ tokens.
The producer generates subsequent tokens from left to right using a local language model.
At position $i$, the current text prefix determines the $h$-token context $c_i$.
The producer ranks the $k$ highest-probability tokens under the current prefix and uses the detection interface to infer the color of each corresponding pair in order.
It selects the first green candidate, or the highest-ranked candidate if all $k$ are red.
The selected token is appended to the current text and is not revised thereafter.
The construction continues until the desired text length is reached.
Let $n$ denote the number of unique context-token pairs induced by the completed text.
This construction applies to Green-Red watermark detection interfaces with the following three properties.
\begin{enumerate}[label=(D\arabic*), noitemsep, leftmargin=*]
  \item For every previously unqueried context-token pair, its color remains $\mathrm{Bernoulli}(\gamma)$ conditioned on all preceding queries and responses.
  \item The requester can choose context-token pairs adaptively and infer the exact color of each queried pair before choosing its next query.
  \item The detection rule is public and returns a positive decision once the green pair count $g$ reaches a threshold determined by $(n,\gamma,\alpha)$.
\end{enumerate}

Condition (D2) concerns what can be inferred across adaptive queries.
For example, an aggregate interface satisfies it if comparing responses to queries that differ by one pair reveals that pair's color.
VOW satisfies (D2) because local finalization reveals the exact color of each queried pair, and it satisfies (D3) through its exact binomial test.
Modeling its PRF as a random function gives (D1).
For the reference calculation, we use the freshness refinement detailed in Appendix~\ref{app:forgery-cost}, which ensures that every queried pair is fresh and every selected pair is unique.
The selected pair is then green exactly when at least one candidate is green, so under (D1) its green probability is $\theta(k)=1-(1-\gamma)^k$.
Let $Z$ be the green count among the $n$ selected pairs, so $Z\sim\mathrm{Binomial}(n,\theta(k))$.
For VOW's exact binomial test, let $g^\star = \min\{g: p(g;n,\gamma)\le\alpha\}$ be the smallest green count that passes the statistical test.
A construction attempt producing $n$ selected pairs succeeds when $Z\ge g^\star$.
Therefore, the success probability of one fixed-$n$ construction attempt is $\Pr[Z\ge g^\star]=\Pr[\mathrm{Binomial}(n,\theta(k))\ge g^\star]$.

We next analyze the cost of the construction.
Suppose each request evaluates one candidate pair and the producer waits to recover its color before sending the next request.
We count each pair evaluation once and each request-response exchange as one adaptive round.
VOW supports this pattern through singleton VOPRF requests.
If another interface requires batching or multiple exchanges to infer one color, these costs may differ.
Let $Q$ be the total number of pair evaluations used to construct the text.
At each of the $n$ positions, the construction evaluates between one and $k$ candidates, so $n\le Q\le kn$.
Since candidate $j$ is evaluated only when the preceding $j-1$ candidates are red, the expected number of construction-phase evaluations per selected pair is
\begin{equation}
  \frac{\mathbb{E}[Q]}{n}
  =\frac{\theta(k)}{\gamma}.
\end{equation}
The ratio $Q/n$ measures the number of construction-phase pair evaluations per unique pair counted in the completed text.
Alternatively, querying all $k$ candidates in one batch at each position uses $kn$ evaluations and $n$ adaptive rounds, whereas sequential early stopping uses $Q$ evaluations and $Q$ adaptive rounds.
Both variants use $\Theta(n)$ adaptive rounds for fixed $k$ because the candidates at the next position depend on the token selected at the current one.
By comparison, an honest audit evaluates all $n$ pairs in one non-adaptive batch.
Deeper batching can reduce the number of adaptive rounds by evaluating a candidate tree, at a cost of exponentially more pair evaluations.
We do not analyze this trade-off further.

The producer can determine from the queried colors whether the completed text will receive a positive decision, so $Q$ counts only the construction cost of one attempt.
Under VOW's batched transcript format, obtaining a transferable record for a successful text requires a separate ordinary audit of its final pair set.
For the sequential realization, this adds $n$ evaluations and one round, giving $Q+n$ evaluations and $Q+1$ rounds in total.
Section~\ref{sec:evaluation-unforgeability} evaluates the construction on concrete Green-Red detectors.

\subsection{Evasion}
\label{sec:analysis-evasion}

An adversary evades detection by attempting to remove the watermark from content it already holds.
We treat this robustness question as empirical, as stated in Section~\ref{sec:tm-scope}.
The $\mathrm{IsGreen}$ predicate depends on an exact $(h+1)$-token sequence, so deletion, insertion, or replacement changes the PRF inputs of every detection window spanning the edited position.
The affected windows yield fresh predicate values, which dilutes the signal.
A larger $h$ widens each window and increases the number of positions a single edit disturbs, which is the attack amplification effect observed for Green-Red watermarks~\cite{kirchenbauer2023watermark}.
Section~\ref{sec:evaluation-robustness} quantifies this behavior and bounds the setting to largely unedited content.

\section{Evaluation}
\label{sec:evaluation}

We evaluate VOW along four dimensions: detectability and text quality, insertion and detection overhead, robustness to removal attacks, and the cost and success of oracle-assisted text construction, then conclude with a qualitative comparison against the baselines.

\subsection{Experimental Setup}

\noindent\textbf{Baselines and watermark parameters.}
We compare VOW with five representative watermarking schemes:
\begin{itemize}[noitemsep, leftmargin=*]
  \item LeftHash \& SelfHash (KGW)~\cite{kirchenbauer2024reliability}: LeftHash uses one preceding token as context ($h=1$), while SelfHash uses four preceding tokens ($h=4$) and includes the candidate token in the PRF input.
  \item RDF~\cite{kuditipudi2024robust}: A distortion-free watermark designed for robustness, whose probabilistic detection procedure is much more computationally expensive than those of the other baselines.
  \item PDW~\cite{fairoze2025publicly}: A publicly detectable watermark that embeds a cryptographic signature into the generated text. Because the signature requires approximately \num{1000} tokens, PDW targets long-text settings.
  \item UPV~\cite{liu2024unforgeable}: A learning-based publicly verifiable watermark that uses two separate neural networks for vocabulary partitioning and public detection.
\end{itemize}
We use the recommended settings for each baseline.
For VOW, we fix $h=2$ and set $\gamma=0.5$ and $\delta=3.0$ by a parameter grid search detailed in Appendix~\ref{app:parameter-selection}.

\noindent\textbf{Detection settings.}
For detectors with a statistical significance parameter,
we set the nominal level $\alpha$ to $10^{-5}$ for VOW, LeftHash, and SelfHash, and to $10^{-2}$ for RDF due to its high detection cost.
UPV uses its default classifier threshold of $0.5$, while PDW provides a binary decision; neither is calibrated to a nominal false positive rate (FPR).

\noindent\textbf{Datasets and models.}
We use Qwen2.5-3B~\cite{qwen25} for text generation and Qwen2.5-7B~\cite{qwen25} for perplexity evaluation.
For open-ended generation, prompts are drawn from C4~\cite{raffel2020exploring}.
For the robustness and fixed-$k$ construction experiments, we use ELI5~\cite{fan2019eli5} prompts and generate responses with Qwen2.5-3B-Instruct.
More data preparation details are \mbox{provided in Appendix~\ref{app:data-preparation}}.

\subsection{Detectability vs. Quality}
\label{sec:evaluation-performance}

We first evaluate the detection accuracy of VOW on short texts and the effects on text quality and downstream utility.

\noindent\textbf{Detectability on short texts.}
A practical watermark should be detectable from short outputs.
We measure the true positive rate (TPR) as the number of tokens increases, following the detection settings described above.
\begin{figure}[t]
  \centering
  \includegraphics{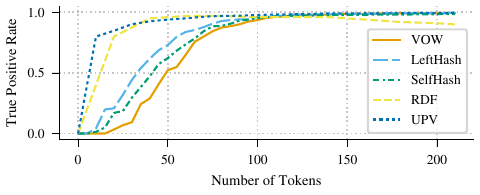}
  \caption{True positive rate (TPR) of different watermarking schemes at varying token lengths.}
  \label{fig:tpr-token-num}
\end{figure}

At the common nominal level $\alpha=10^{-5}$, Figure~\ref{fig:tpr-token-num} shows that VOW's TPR rises steadily with text length and becomes comparable to LeftHash and SelfHash once roughly 100 tokens are available.
Its TPR is 0.936 at 100 tokens and 0.994 at 200 tokens.
RDF's curve rises quickly under its relaxed nominal level of $10^{-2}$, which is \num{1000} times less strict, but its TPR is non-monotonic, dropping from 0.925 at 80 tokens to 0.88 at 200 tokens.
UPV also rises quickly, but its default classifier threshold is not calibrated to a nominal FPR.
PDW is designed for longer texts and thus is not included in Figure~\ref{fig:tpr-token-num}.
We separately evaluate PDW on \num{1000}-token texts, where it reaches a TPR of 0.974.

\begin{table}[t]
  \centering
  \caption{False-positive counts on unwatermarked C4 texts, with outward-rounded 95\% exact Poisson confidence intervals in brackets.
  Counts are based on $10^6$ texts, except RDF, which uses $10^5$ texts.}
  \label{tab:fpr-calibration}
  \small
  \setlength{\tabcolsep}{2pt}
  \newcommand{\fpci}[2]{{\tiny\color{black!60}[\num{#1}--\num{#2}]}}
  \begin{tabular}{c@{\hspace{6pt}}r@{\,}l@{\hspace{6pt}}r@{\,}l@{\hspace{6pt}}r@{\,}l@{\hspace{6pt}}r@{\,}l}
    \toprule
    $\alpha$ & \multicolumn{2}{c}{VOW} & \multicolumn{2}{c}{LeftHash} & \multicolumn{2}{c}{SelfHash} & \multicolumn{2}{c}{RDF} \\
    \midrule
    $10^{-2}$ & \num{8113} & \fpci{7937}{8292} & \num{10939} & \fpci{10734}{11146} & \num{7700} & \fpci{7528}{7874} & \num{248} & \fpci{218}{281} \\
    $10^{-3}$ & \num{747} & \fpci{694}{803} & \num{1218} & \fpci{1150}{1289} & \num{1514} & \fpci{1438}{1593} & \multicolumn{2}{c}{--} \\
    $10^{-4}$ & \num{75} & \fpci{58}{95} & \num{139} & \fpci{116}{165} & \num{409} & \fpci{370}{451} & \multicolumn{2}{c}{--} \\
    $10^{-5}$ & \num{6} & \fpci{2}{14} & \num{10} & \fpci{4}{19} & \num{136} & \fpci{114}{161} & \multicolumn{2}{c}{--} \\
    \bottomrule
  \end{tabular}
\end{table}

\noindent\textbf{False-positive calibration.}
Table~\ref{tab:fpr-calibration} compares the detectors' observed false-positive counts on unwatermarked texts with the reference counts obtained by multiplying the sample size by the nominal significance level.
At $\alpha=10^{-5}$, VOW and LeftHash produce 6 and 10 false positives, respectively, compared with the nominal reference count of 10 in \num{1000000} samples.
VOW's observed counts are below the nominal references at all four significance levels, while LeftHash shows modest FPR inflation at $10^{-2}$ through $10^{-4}$.
SelfHash is increasingly miscalibrated in the lower tail: at $\alpha=10^{-5}$, it produces 136 false positives, or 13.6 times the nominal rate.
RDF produces 248 false positives in \num{100000} samples at $\alpha=10^{-2}$, corresponding to an empirical FPR of \num{2.48e-3}.
PDW and UPV do not provide $\alpha$-indexed detectors and are therefore not included in the table.
At its default threshold of 0.5, UPV's public detection network flags 1748 of \num{1000000} unwatermarked texts, for an empirical FPR of $1.748\times 10^{-3}$ (detailed in Appendix~\ref{app:upv-detection-reliability}).

\begin{figure}[t]
  \centering
  \includegraphics{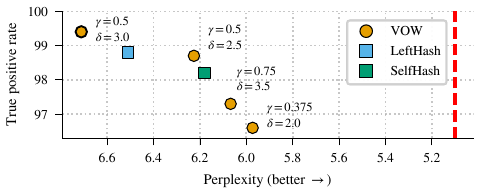}
  \caption{Trade-off between true positive rate and perplexity (PPL) for different watermarking schemes. The red dashed line represents PPL of non-watermarked~texts.}
  \label{fig:effectiveness}
\end{figure}

\noindent\textbf{Quality-detectability trade-off.}
We evaluate the quality cost of watermarking with perplexity (PPL), where lower values indicate more fluent and natural text.
Figure~\ref{fig:effectiveness} plots TPR against PPL for samples generated with multinomial sampling at temperature 0.7.
With the default setting $(\gamma,\delta)=(0.5,3.0)$, VOW reaches a TPR of 0.994 at a PPL of 6.71, compared with 5.10 without watermarking.
Reducing $\delta$ to 2.5 lowers PPL to 6.23 while retaining a TPR of 0.987, placing VOW close to the KGW variants in the quality--detectability trade-off.
RDF, UPV, and PDW are omitted from the figure for readability because their perplexity scores are much larger: 23.68, 25.46, and 1885.12, respectively.

\noindent\textbf{Utility on downstream tasks.}
Perplexity does not fully capture the cost of watermarking on LLM performance.
We therefore evaluate mathematical reasoning on GSM8K~\cite{cobbe2021training} with 4-shot prompting and code generation on HumanEval~\cite{chen2021codex} with 0-shot prompting, using Qwen2.5-3B-Instruct.
These low-entropy tasks are challenging for watermarking because small token-level biases can affect precise outputs.
\begin{table}[t]
  \centering
  \caption{Impact of watermarking on downstream task performance: accuracy (\%) on GSM8K (4-shot) and pass@1 (\%) on HumanEval (0-shot).}
  \label{tab:downstream-tasks}
  \small
  \begin{tabular}{lSS}
    \toprule
    Method    & {GSM8K (4-shot)} & {HumanEval (0-shot)} \\
    \midrule
    No watermark       & 81.12            & 54.88                \\
    VOW ($\delta=3.0$) & 78.01            & 50.00                \\
    VOW ($\delta=2.5$) & 78.54            & 53.66                \\
    LeftHash           & 75.59            & 45.73                \\
    SelfHash           & 74.37            & 46.95                \\
    RDF                & 2.43             & 3.66                 \\
    PDW                & 1.36             & 0.0                  \\
    UPV                & 78.95            & 50.00                \\
    \bottomrule
  \end{tabular}
\end{table}
As shown in Table~\ref{tab:downstream-tasks}, all watermarking methods reduce downstream performance, but the magnitude differs substantially.
RDF and PDW fall below 4\% on both metrics.
VOW, KGW variants, and UPV retain much higher utility, with VOW outperforming LeftHash and SelfHash on both tasks.
Reducing $\delta$ from 3.0 to 2.5 improves VOW's downstream performance while slightly reducing detectability, illustrating the trade-off between detection strength and utility.

\subsection{Efficiency and Overhead}
\label{sec:evaluation-overhead}

We measure watermark insertion and detection costs.
Evaluated methods include both optimized and unoptimized VOW to isolate the effect of the optimizations in Section~\ref{sec:method-optimizations}.
All measurements use an AMD Ryzen 9 7900X CPU and an NVIDIA RTX 5070 Ti GPU.

\noindent\textbf{Computational overhead.}
We report generation latency and detection runtime, focusing on inter-token latency (ITL), the average time to generate each token after the first.
Most schemes have a time to first token (TTFT) comparable to the unwatermarked baseline, except RDF, whose precomputation adds approximately 30 seconds of latency.

\begin{table}[t]
  \centering
  \caption{Average inter-token latency (ITL) and detection runtime for different watermarking schemes (in milliseconds). Detection runtime excludes network round-trip time.}
  \label{tab:watermark-overhead}
  \small
  \setlength{\tabcolsep}{4.5pt}
  \begin{tabular}{
      l
      S[table-format=3.2]
      S[table-format=4.2]
      S[table-format=5.2]
      S[table-format=5.2]
    }
    \toprule
    \multirow{2}{*}{Method} & \multicolumn{3}{c}{ITL (ms) by Batch Size}  & \multicolumn{1}{c}{\multirow{2}{*}{\makecell{Detection\\ Time (ms)}}} \\
    \cmidrule(lr){2-4}
    & {1}                              & {8}       & {32}    &  \\
    \midrule
    No watermark            & 25.42                          & 27.19   & 32.95     & {---} \\
    \midrule
    VOW (unoptimized)       & 429.18                         & 3280.86 & 12890.53  & {\multirow{4}{*}{\tablenum[table-format=5.2]{62.20}}}  \\
    VOW (multinomial)       & 31.10                          & 50.81   & 79.77     &  \\
    VOW (top-10)            & 36.41                          & 43.63   & 59.94     &  \\
    VOW (top-50)            & 43.52                          & 48.35   & 87.78     &  \\
    \midrule
    LeftHash                & 27.25                          & 34.71   & 41.20     & 118.18 \\
    SelfHash                & 108.21                         & 714.43  & 2774.97   & 265.06 \\
    RDF                     & 21.71                          & 27.64   & 33.95     & 24075.92 \\
    PDW                     & 180.83                         & {---}   & {---}     & 584.20 \\
    UPV                     & 360.65                         & 2967.60 & 11103.28  & 7.73   \\
    \bottomrule
  \end{tabular}
\end{table}

Table~\ref{tab:watermark-overhead} reports ITL at different batch sizes and detection times for 500-token texts.
Unoptimized VOW is expensive because it scans the whole vocabulary, as do SelfHash and UPV.
The optimized VOW variants avoid this full scan and reduce ITL by over 99.3\% at batch size 32 relative to unoptimized VOW\@.
Counting both TTFT and ITL, LeftHash has the lowest generation overhead among evaluated watermarking methods, while optimized VOW remains within a practical range.
RDF's ITL alone is the lowest in Table~\ref{tab:watermark-overhead}, but the roughly 30-second TTFT noted above makes its total generation cost far higher, limiting its use in real-time generation.
PDW has high ITL because its rejection sampling rejects about 75\% of candidates and does not support batch generation in its original implementation.

Most methods finish detection within several hundred milliseconds on 500-token texts.
RDF is the main exception, requiring over 24 seconds under our evaluation settings.
UPV is fastest on our platform because its neural network detector uses GPU inference.
VOW's measured detection computation takes 62.20 ms; end-to-end latency additionally depends on the deployment network.

\noindent\textbf{Communication overhead.}
VOW communicates blinded inputs, evaluated outputs, and a batched proof during detection.
In our implementation, the total transfer scales linearly with input length and is approximately 13.5 times the plaintext size.
More details are provided in Appendix~\ref{app:communication-overhead}.
The KGW variants and RDF likewise require an interactive detection service when their keys are held by the provider.
PDW and UPV support local detection, so their communication is limited to one-time distribution of public verification material or a public detection network.

\begin{figure*}[t]
  \centering
  \includegraphics{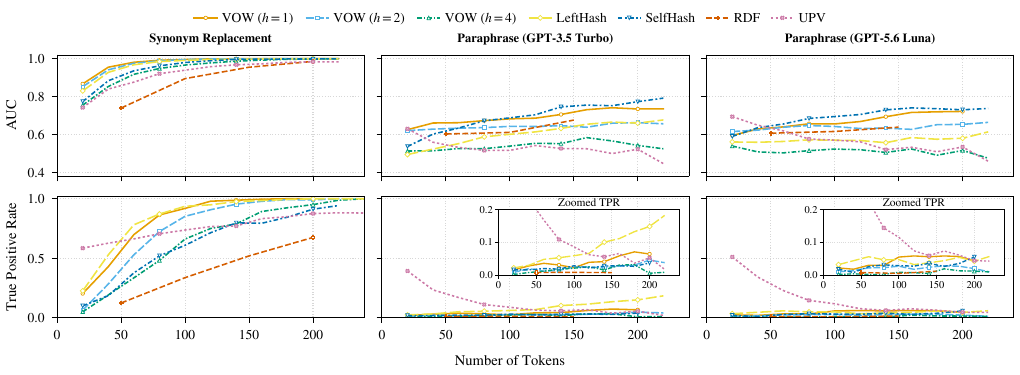}
  \caption{AUC (top) and TPR (bottom) of different watermarking schemes under synonym replacement and paraphrasing attacks, with varying numbers of tokens used for detection.}
  \label{fig:robustness-auc-tpr}
\end{figure*}
\subsection{Robustness Against Removal Attacks}
\label{sec:evaluation-robustness}

We evaluate robustness against two common removal attacks~\cite{zhao2025sok}: synonym replacement and paraphrasing.

\noindent\textbf{Metrics.}
We use two complementary metrics.
The area under the ROC curve (AUC) measures separability between attacked watermarked samples and non-watermarked samples, and is commonly used in prior robustness evaluations~\cite{kirchenbauer2024reliability}.
Because high AUC does not by itself imply practical detection at low FPRs, we also report TPR at a nominal significance of $10^{-2}$ when applicable.
Metrics are computed on 500 attacked watermarked samples and 500 non-watermarked samples.

\noindent\textbf{Attack model.}
We evaluate a synonym replacement attack and two paraphrasing attacks using different LLM backends.
\begin{enumerate}[noitemsep, leftmargin=18pt]
  \item {Synonym replacement}: The attack replaces 30\% of eligible words (nouns, verbs, etc.) with synonyms sampled from the top-15 predictions of DistilBERT~\cite{sanh2019distilbert}.
  \item {Paraphrase by GPT-3.5 Turbo}: We use the paraphrasing setup from prior work~\cite{kirchenbauer2024reliability} to rewrite the watermarked text while preserving its meaning.
  \item {Paraphrase by GPT-5.6 Luna}: We repeat the same paraphrasing prompt with GPT-5.6 Luna to evaluate a stronger rewriting attack.
\end{enumerate}

The attacked texts retain substantial semantic similarity overall, with lower values for PDW under paraphrasing.
Paraphrasing also approximately preserves length except that PDW's longer source texts are substantially compressed (detailed in Appendix~\ref{app:data-preparation}).

\noindent\textbf{Results.}
Figure~\ref{fig:robustness-auc-tpr} reports AUC and TPR under synonym replacement and paraphrasing.
Under synonym replacement, most schemes reach AUC above 0.95 with sufficient tokens, but TPR reveals important differences.
RDF is the clear exception, remaining below 0.7 TPR even at the longest token counts evaluated.
PDW fails under both synonym replacement and paraphrasing because these edits disrupt its cryptographic signature, so it is omitted from the figure for readability.
The default VOW setting with $h=2$, the KGW variants, and UPV all reach TPR above 0.85 with enough tokens.
The VOW variants also show the expected context-length trend: $h=1$ retains more signal after editing, whereas $h=4$ retains less, consistent with Section~\ref{sec:analysis-evasion}.

Paraphrasing is substantially more damaging than synonym replacement.
Against GPT-3.5 Turbo, the KGW variants obtain lower AUC than reported in prior work~\cite{kirchenbauer2024reliability} under our generation and attack settings.
At longer detection lengths, VOW's default $h=2$ setting generally matches or exceeds LeftHash and UPV in AUC, while $h=1$ retains more separation and $h=4$ retains less.
Nevertheless, the TPR of every VOW context length remains below 0.1 under both paraphrasing attacks.
Under GPT-5.6 Luna paraphrasing, UPV's TPR peaks at short detection lengths and approaches zero as more tokens are included.

\begin{figure}[t]
  \centering
  \includegraphics{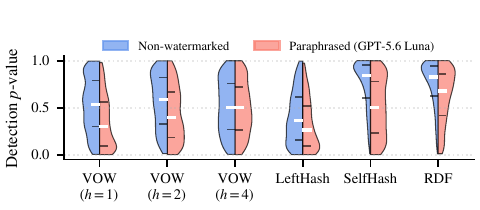}
  \caption{Distribution of detection $p$-values for unwatermarked samples and watermarked texts after paraphrasing by GPT-5.6 Luna.}
  \label{fig:p-vals-gpt5}
\end{figure}

Figure~\ref{fig:p-vals-gpt5} compares unwatermarked samples with watermarked texts after GPT-5.6 Luna paraphrasing.
The two distributions overlap substantially for every scheme shown, consistent with the low TPRs in Figure~\ref{fig:robustness-auc-tpr}.
For VOW, this overlap increases with context length: $h=1$ retains the clearest separation, the default $h=2$ retains less, and $h=4$ retains almost none.
No evaluated scheme remains reliably detectable after paraphrasing.
These results bound watermark-based FPA to relatively intact content: a positive VOW result indicates consistency with the committed watermark key, not robustness to subsequent rewriting.

\subsection{Evaluation of the Fixed-\texorpdfstring{$k$}{k} Construction}
\label{sec:evaluation-unforgeability}

We evaluate the concrete fixed-$k$ construction presented in Section~\ref{sec:analysis-forgery}.
The producer generates from unseen ELI5 prompts with Qwen2.5-3B-Instruct and learns candidate colors only through the detection protocol, without access to $\mathsf{wk}$.
At each position, it submits one candidate pair per request, up to a budget of $k$, and waits for each color before deciding whether to query the next candidate.
We generate \num{1000} texts per configuration, each with $n=300$ unique context-token pairs, and test them at nominal significance $\alpha=10^{-5}$.

For fresh candidate pairs under (D1), the selected-pair green probability is $\theta(k)=1-(1-\gamma)^k$, and the expected construction cost per final pair is $\mathbb{E}[Q]/n=\theta(k)/\gamma$.
We report the fraction of constructed texts receiving a positive decision (attack success rate, ASR), measured green rate, perplexity, and $Q/n$, where $Q$ is the number of pair evaluations used during construction.
Because this execution sends one pair per request, $Q$ also equals the number of adaptive construction rounds and is $\Theta(n)$ for fixed $k$.
The reported construction costs exclude network latency and, for VOW, VOPRF proof-generation time.
Under VOW's current transcript format, obtaining a transferable record for a successful text requires a subsequent ordinary audit of the $n$ final pairs, giving total costs of $Q+n$ evaluations and $Q+1$ rounds.

\begin{table}[t]
  \centering
  \caption{Our fixed-$k$ construction against VOW with $n=300$ unique context-token pairs, $\gamma=0.5$, and $\alpha=10^{-5}$.
  The reference predictions under (D1) are $\theta(k)=1-(1-\gamma)^{k}$ and $\theta(k)/\gamma$.}
  \label{tab:forgery-k}
  \small
  \setlength{\tabcolsep}{4.1pt}
  \begin{tabular}{lS[table-format=1.3]S[table-format=1.3]S[table-format=1.3]S[table-format=1.3]S[table-format=3.0]S[table-format=2.2]}
    \toprule
    $k$ & {Pred.\ $\theta(k)$} & {Green rate} & {Pred.\ $Q/n$} & {$Q/n$} & {ASR (\%)} & {PPL} \\
    \midrule
    1 & 0.500 & 0.494 & 1.000 & 1.000 & 0 & 3.412 \\
    2 & 0.750 & 0.749 & 1.500 & 1.502 & 100 & 5.732 \\
    3 & 0.875 & 0.874 & 1.750 & 1.749 & 100 & 7.778 \\
    4 & 0.938 & 0.936 & 1.875 & 1.878 & 100 & 9.312 \\
    5 & 0.969 & 0.967 & 1.938 & 1.943 & 100 & 10.323 \\
    \bottomrule
  \end{tabular}
\end{table}

At $k=1$ the adversary observes a color but has nothing to substitute, so the green rate stays at $\gamma$ and no sample passes.
At $k=2$ all samples pass, with a mean green rate of $0.749$.
Their perplexity is 5.732, compared with 5.246 for ordinarily watermarked texts.
Larger values of $k$ increase the green rate and construction cost while degrading text quality, although $k=2$ already achieves 100\% ASR in this experiment.

\begin{table}[t]
  \centering
  \caption{Our fixed-$k$ construction evaluated on three Green-Red schemes with $n=300$ unique context-token pairs and $k=2$, each at its own $\gamma$ and at $\alpha=10^{-5}$.}
  \label{tab:forgery-baselines}
  \small
  \begin{tabular}{lS[table-format=1.2]S[table-format=1.3]S[table-format=1.3]S[table-format=3.1]}
    \toprule
    Scheme & {$\gamma$} & {Pred.\ $Q/n$} & {$Q/n$} & {ASR (\%)} \\
    \midrule
    VOW      & 0.50 & 1.500 & 1.502 & 100 \\
    LeftHash & 0.25 & 1.750 & 1.745 & 100 \\
    SelfHash & 0.25 & 1.750 & 1.765 & 90.2 \\
    \bottomrule
  \end{tabular}
\end{table}

Before applying the construction, each detector achieves a TPR above 0.99 on ordinarily watermarked texts with $n=300$ unique context-token pairs.
The reference predictions use the fresh-pair Bernoulli condition in (D1).
LeftHash samples candidates without replacement from a fixed-size green list, while SelfHash uses candidate-dependent seeding, so their concrete color responses need not follow the independent Bernoulli model exactly.
For these two baselines, the Bernoulli calculation is therefore a reference prediction rather than a scheme-level guarantee.
Nevertheless, the measured $Q/n$ values remain close to the predictions.
At $k=2$, all \num{1000} constructed VOW and LeftHash samples passed, while 902 of the \num{1000} SelfHash samples passed.
The construction therefore succeeds on all three evaluated interfaces, with deviations from the Bernoulli reference reflected in the realized ASR.

\begin{table*}[t]
  \centering
  \caption{Qualitative comparison of audit-interface properties, core performance, and overhead.}
  \label{tab:evaluation-summary}
  \small

  \begin{threeparttable}

    \begin{tabular}{l|ccc|cccc|cc}
      \toprule
      \multirow{2}{*}{Method}
      & \multicolumn{3}{c|}{Audit-Interface Properties}
      & \multicolumn{4}{c|}{Core Performance}
      & \multicolumn{2}{c}{Practicality \& Overhead} \\
      \cmidrule(lr){2-4} \cmidrule(lr){5-8} \cmidrule(lr){9-10}
      & \makecell{Input\\Privacy}
      & \makecell{Result\\Verifiability}
      & \makecell{Transferable\\Evidence}
      & \makecell{Detectability\\(Short Text)}
      & \makecell{Text\\Quality}
      & \makecell{Robustness\\(Synonym)}
      & \makecell{Robustness\\(Paraphrasing)}
      & \makecell{Insertion\\Overhead}
      & \makecell{Detection\\Overhead} \\
      \midrule
      VOW      & \CIRCLE & \CIRCLE & \CIRCLE & \CIRCLE & \CIRCLE & \CIRCLE & \Circle & \CIRCLE & \CIRCLE \\
      LeftHash & \Circle & \Circle & \Circle & \CIRCLE & \CIRCLE & \CIRCLE & \Circle & \CIRCLE & \CIRCLE \\
      SelfHash & \Circle & \Circle & \Circle & \CIRCLE & \CIRCLE & \CIRCLE & \Circle & \Circle & \CIRCLE \\
      RDF      & \Circle & \Circle & \Circle & \CIRCLE & \LEFTcircle & \LEFTcircle & \Circle & \LEFTcircle & \Circle \\
      PDW      & \CIRCLE & \CIRCLE & \CIRCLE & \Circle & \Circle & \Circle & \Circle & \LEFTcircle & \CIRCLE \\
      UPV      & \CIRCLE & \LEFTcircle & \LEFTcircle & \CIRCLE & \LEFTcircle & \CIRCLE & \Circle & \Circle & \CIRCLE \\
      \bottomrule
    \end{tabular}

    \begin{tablenotes}[flushleft]
      \footnotesize
    \item[] \textit{Note:}
      ${\normalfont\CIRCLE}$ denotes explicit support or strong performance;
      ${\normalfont\LEFTcircle}$ denotes partial support, a restricted guarantee, or empirical/heuristic evidence;
      ${\normalfont\Circle}$ denotes no support or poor performance.
      Input privacy concerns the submitted content apart from query count and transport metadata.
      Result verifiability asks whether the detection result can be independently checked against the declared marking setup.
      Transferable evidence asks whether a later reviewer given the text and retained record can independently re-check that result.

    \item[]
      The robustness columns summarize the 30\% synonym-replacement attack and the two LLM paraphrasing attacks in Section~\ref{sec:evaluation-robustness}.
      No evaluated scheme remains reliably detectable after paraphrasing.
    \end{tablenotes}

  \end{threeparttable}
\end{table*}

\subsection{Summary}
\label{sec:evaluation-summary}

Table~\ref{tab:evaluation-summary} compares input privacy, result verifiability, transferable evidence, core performance, and overhead.
Under provider-held keys, the KGW variants and RDF require remote detection, so the auditor must disclose the submitted text and cannot independently verify or later re-check the result.
The KGW variants nevertheless provide strong short-text detectability and text quality at low cost, whereas RDF has an expensive permutation test and substantially higher perplexity in our evaluation.
PDW and UPV avoid interaction through local public detection.
PDW meets all three audit-interface goals, but its cryptographic payload requires long outputs and substantially reduces text quality.
UPV keeps the input local and allows a third party to reproduce the detector output, but its released detector is not cryptographically bound to the marking procedure and provides no false-positive control at a specified nominal level.
VOW instead targets interactive detection on short outputs.
It hides the input from the key holder, binds accepted results to a published setup record, and leaves transferable evidence, while retaining short-text detectability, competitive text quality, and practical overhead.

With sufficient tokens, VOW, the KGW variants, and UPV retain useful detectability under synonym replacement, whereas RDF is weaker and PDW's cryptographic payload is disrupted.
No evaluated scheme remains reliably detectable after LLM paraphrasing.
The fixed-$k$ construction produces detector-positive text for VOW, LeftHash, and SelfHash when the requester can infer the exact color of each queried pair.

\section{Related Work}
\label{sec:related-work}

We discuss VOW in relation to watermarking schemes for LLM outputs, efforts to reduce trust in the detector, attacks on detection interfaces, and the auditing of LLM services.
Model and dataset watermarking~\cite{zhao2023protecting,zhang2024remark,guo2023domain,cui2025robust} are adjacent topics that we do not cover.

\subsection{Watermarking Schemes for LLM Outputs}

A large body of work builds on the logit-biasing paradigm introduced by KGW~\cite{kirchenbauer2023watermark,kirchenbauer2024reliability}, in which a secret key determines which tokens are favored during generation.
Two goals have received most attention.
The first is robustness to modification, pursued through sentence-level and semantic representations designed to survive paraphrasing~\cite{hou2024semstamp,hou2024ksemstamp,ren2024robust} and through formal analysis of restricted variants, such as the static green-list scheme analyzed by Zhao et al.~\cite{zhao2024provable}.
The second is preserving generation quality.
Adaptive watermarking~\cite{liu2024adaptive} and dual-key designs~\cite{zhu2024duwak} adjust watermark strength or key usage to this end.
DiPmark~\cite{wu2024resilient} preserves the output distribution while improving robustness under its detection assumptions.
RDF~\cite{kuditipudi2024robust} samples without distortion and serves as a baseline in our evaluation, while other work studies cryptographic undetectability under stated assumptions~\cite{christ2024undetectable}.

These schemes assume a trusted detection path.
A keyed detector must receive the submitted text and report the outcome honestly, and the schemes above take both for granted.
VOW leaves the insertion rule in this family and changes the detection interface.

\subsection{Reducing Trust in Detection}

Publicly detectable watermarks such as PDW~\cite{fairoze2025publicly} and related cryptographic constructions~\cite{christ2024pseudorandom} let anyone verify a watermark with public information, so the text need not be disclosed for detection.
These schemes embed signatures or codewords as verification payloads and therefore require text long enough to carry them, limiting their use on short outputs.

Earlier multimedia zero-knowledge detectors conceal watermark secrets and intermediate detector values to limit oracle leakage while proving presence~\cite{adelsbach2001zk,adelsbach2003disclosure}.
PVMark similarly proves correct keyed detection~\cite{duan2026pvmark}.
Both treat the examined asset as a shared or public proof input, whereas VOW hides the audited text from the detection service.

Learning-based schemes take a different route.
UPV~\cite{liu2024unforgeable} trains and releases a detection neural network, allowing an auditor to run detection locally without disclosing the content.
Verification is then reproducible by a third party, which VOW also provides.
The difference is the reference for that check.
VOW binds a transcript to the key and detector specification committed before use, whereas re-running a released network establishes only what that network outputs, without tying the result to such a prior commitment.
The learned detector also provides no false-positive control at a specified nominal level.
VOW instead retains interaction with the key holder while supporting input-private, verifiable detection on short outputs.

\subsection{Attacks on Detection Interfaces}

Detector-oracle attacks long predate LLM watermarking.
Linnartz and van Dijk analyzed leakage from repeated binary detector decisions, while Kalker et al. estimated a secret watermark pattern from black-box behavior~\cite{linnartz1998sensitivity,kalker1998detector}.
Recent LLM work recovers watermarking rules from model queries~\cite{jovanovic2024stealing}.
Pang et al. use aggregate detector scores for removal and spoofing and explicitly connect their attack to this classical line~\cite{pang2024nofreelunch}.

Our oracle-assisted text construction instantiates the long-standing detector-oracle concern for Green-Red watermark detection when the interface allows the exact color of each queried pair to be inferred.
Prior defenses either harden the detector's decision boundary or identify related plaintext queries~\cite{troncoso2006sensitivity,barni2014smart}.
VOW's blinding prevents the detection service from applying the latter defense, exposing a tension between input privacy and content-aware oracle monitoring.

\subsection{Auditing of LLM Services}

Work on auditing LLM services studies functional consistency~\cite{amirizaniani2024auditllm}, data provenance~\cite{wu2025synthetic}, service integrity risks such as model downgrading~\cite{zhu2025auditing,cai2025are}, and system-level privacy risks such as prompt caching side channels~\cite{gu2025auditing}.
Methods for detecting model downgrading infer service behavior from output distributions, using log-probabilities~\cite{cai2025are} or rank-based uniformity tests~\cite{zhu2025auditing}, or rely on trusted execution environments~\cite{cai2025are}.
These methods examine provider conduct from a customer's perspective, whereas first-party provenance auditing asks whether selected artifacts already in the auditing organization's custody carry the signal of a declared marked generation pipeline.
The two settings are complementary.

\section{Limitations}
\label{sec:limitations}

\noindent\textbf{Interpreting a detection record.}
VOW makes the prescribed detection result for a presented text under a selected setup record re-checkable, but the record does not identify the author or establish how the text was generated.
A positive result supports association with the declared generation pipeline only when use of the watermarking key is confined to that pipeline and its prescribed detection service, and the text was not assembled using detector feedback.
The nominal false-positive rate applies only under the stated null model of unwatermarked text selected independently of the committed key and detection oracle.
Whether the provider faithfully applies its marking policy is a separate question that requires sampling covered outputs.

\noindent\textbf{Artifact and setup selection.}
Transferable evidence covers all detector inputs induced by the presented text under one selected setup record.
It does not establish that the text is the complete artifact or that the applicable setup record was selected before testing.
Submitting a short excerpt can reduce detector sensitivity and make a negative result more likely, while testing several valid setup records and reporting only a passing result creates a multiple-testing problem.
The artifact boundary and applicable setup record must therefore be fixed by the audit policy before detection.

\noindent\textbf{Robustness to modification.}
Paraphrasing substantially reduces detectability for every scheme we evaluate, including VOW.
VOW is therefore best suited to relatively intact content, and a negative result after substantial rewriting says little about whether the original text carried the watermark.

\noindent\textbf{Detector structure.}
VOW currently targets Green-Red detectors built from pointwise PRF evaluations.
Other structures require different oblivious and verifiable mechanisms.

\noindent\textbf{Interaction and leakage.}
VOW requires interaction with a detection service, and the service can refuse or abort an audit.
Its communication grows linearly with the number of unique context-token pairs and, in our implementation, is approximately $13.5\times$ the plaintext size.
Input privacy hides the queried pairs but reveals their count and does not cover transport metadata such as account identity, timing, and request boundaries.
Arbitration additionally requires the auditor to disclose the text to the chosen reviewer.

\noindent\textbf{Attestation when metadata survives.}
Where signed provenance metadata remains associated with content, attestation provides a more direct link to the generation event than watermarking.
Content provenance standards such as C2PA take this approach for media files~\cite{c2pa}.
California likewise requires provenance information for covered image, video, and audio content~\cite{ca_sb942}.
VOW addresses relatively intact text that has been separated from such external metadata.

\section{Conclusion}
\label{sec:conclusion}

When a provider retains the watermark key, a first-party provenance audit ordinarily requires the organization to disclose its text to the detection service and trust the answer.
VOW uses a VOPRF to hide detector inputs and bind returned outputs to the watermark key committed in a signed setup record.
The auditor performs the statistical test locally, so the service can refuse but cannot make an incorrect result pass verification.
A reviewer holding the text and setup record can later reconstruct the prescribed result from the retained transcript without the watermarking key or further provider interaction.
At $\alpha=10^{-5}$, VOW produced six false positives among one million unwatermarked texts and reached a TPR of 0.994 on watermarked texts of 200 tokens, with competitive text quality and practical overhead.

Our fixed-$k$ construction uses detector feedback to assemble text that passes detection without the watermarking key.
Paraphrasing substantially weakens every evaluated scheme, so FPA using watermarks is best suited to relatively intact content.
Within this scope, VOW supports audits of short text that keep detector inputs private and results independently checkable while the provider retains the watermark key.

\bibliographystyle{ACM-Reference-Format}
\bibliography{references}

@misc{rfc9497,
  series       = {Request for Comments},
  number       = 9497,
  howpublished = {RFC 9497},
  publisher    = {RFC Editor},
  doi          = {10.17487/RFC9497},
  url          = {https://www.rfc-editor.org/info/rfc9497},
  author       = {Alex Davidson and Armando Faz-Hernandez and Nick Sullivan and Christopher A. Wood},
  title        = {Oblivious Pseudorandom Functions ({OPRFs}) Using Prime-Order Groups},
  pagetotal    = 61,
  year         = 2023,
  month        = dec
}

@inproceedings{kirchenbauer2023watermark,
  title     = {A Watermark for Large Language Models},
  author    = {Kirchenbauer, John and Geiping, Jonas and Wen, Yuxin and Katz, Jonathan and Miers, Ian and Goldstein, Tom},
  booktitle = {the 40th International Conference on Machine Learning},
  pages     = {17061--17084},
  year      = {2023},
  volume    = {202},
  series    = {Proceedings of Machine Learning Research},
  month     = {23--29 Jul},
  publisher = {PMLR}
}

@inproceedings{kirchenbauer2024reliability,
  title     = {On the Reliability of Watermarks for Large Language Models},
  author    = {John Kirchenbauer and Jonas Geiping and Yuxin Wen and Manli Shu and Khalid Saifullah and Kezhi Kong and Kasun Fernando and Aniruddha Saha and Micah Goldblum and Tom Goldstein},
  booktitle = {The Twelfth International Conference on Learning Representations},
  year      = {2024},
  url       = {https://openreview.net/forum?id=DEJIDCmWOz}
}

@inproceedings{fernandez2023three,
  author    = {Fernandez, Pierre and Chaffin, Antoine and Tit, Karim and Chappelier, Vivien and Furon, Teddy},
  booktitle = {2023 IEEE International Workshop on Information Forensics and Security (WIFS)},
  title     = {Three Bricks to Consolidate Watermarks for Large Language Models},
  year      = {2023},
  volume    = {},
  number    = {},
  pages     = {1-6},
  doi       = {10.1109/WIFS58808.2023.10374576}
}

@article{fairoze2025publicly,
  author    = {Jaiden Fairoze and Sanjam Garg and Somesh Jha and Saeed Mahloujifar and Mohammad Mahmoody and Mingyuan Wang},
  title     = {Publicly-Detectable Watermarking for Language Models},
  volume    = {1},
  number    = {4},
  year      = {2025},
  issn      = {3006-5496},
  doi       = {10.62056/ahmpdkp10},
  journal   = {{IACR} Communications in Cryptology},
  publisher = {International Association for Cryptologic Research}
}

@inproceedings{christ2024pseudorandom,
  author    = {Christ, Miranda
               and Gunn, Sam},
  title     = {Pseudorandom Error-Correcting Codes},
  booktitle = {Advances in Cryptology -- CRYPTO 2024},
  year      = {2024},
  publisher = {Springer Nature Switzerland},
  address   = {Cham},
  pages     = {325--347},
  isbn      = {978-3-031-68391-6}
}

@article{kuditipudi2024robust,
  title   = {Robust Distortion-free Watermarks for Language Models},
  author  = {Rohith Kuditipudi and John Thickstun and Tatsunori Hashimoto and Percy Liang},
  journal = {Transactions on Machine Learning Research},
  issn    = {2835-8856},
  year    = {2024},
  url     = {https://openreview.net/forum?id=FpaCL1MO2C},
  note    = {}
}

@inproceedings{christ2024undetectable,
  title     = {Undetectable Watermarks for Language Models},
  author    = {Christ, Miranda and Gunn, Sam and Zamir, Or},
  booktitle = {Thirty Seventh Conference on Learning Theory},
  pages     = {1125--1139},
  year      = {2024},
  volume    = {247},
  series    = {Proceedings of Machine Learning Research},
  month     = {30 Jun--03 Jul},
  publisher = {PMLR},
  url       = {https://proceedings.mlr.press/v247/christ24a.html}
}

@inproceedings{ren2024robust,
  title     = {A Robust Semantics-based Watermark for Large Language Model against Paraphrasing},
  author    = {Ren, Jie and Xu, Han and Liu, Yiding and Cui, Yingqian and Wang, Shuaiqiang and Yin, Dawei and Tang, Jiliang},
  booktitle = {Findings of the Association for Computational Linguistics: NAACL 2024},
  month     = jun,
  year      = {2024},
  address   = {Mexico City, Mexico},
  publisher = {Association for Computational Linguistics},
  url       = {https://aclanthology.org/2024.findings-naacl.40/},
  doi       = {10.18653/v1/2024.findings-naacl.40},
  pages     = {613--625}
}

@inproceedings{zhang2024remark,
  author    = {Zhang, Ruisi and Hussain, Shehzeen Samarah and Neekhara, Paarth and Koushanfar, Farinaz},
  title     = {{REMARK-LLM}: a robust and efficient watermarking framework for generative large language models},
  year      = {2024},
  isbn      = {978-1-939133-44-1},
  publisher = {USENIX Association},
  address   = {USA},
  booktitle = {the 33rd USENIX Conference on Security Symposium},
  articleno = {102},
  numpages  = {18},
  location  = {Philadelphia, PA, USA},
  series    = {SEC '24}
}

@article{dathathri2024scalable,
  author    = {Dathathri, Sumanth and See, Abigail and Ghaisas, Sumedh and Huang, Po-Sen and McAdam, Rob and Welbl, Johannes and Bachani, Vandana and Kaskasoli, Alex and Stanforth, Robert and Matejovicova, Tatiana and Hayes, Jamie and Vyas, Nidhi and Al Merey, Majd and Brown-Cohen, Jonah and Bunel, Rudy and Balle, Borja and Cemgil, Taylan and Ahmed, Zahra and Stacpoole, Kitty and Shumailov, Ilia and Baetu, Ciprian and Gowal, Sven and Hassabis, Demis and Kohli, Pushmeet},
  title     = {Scalable watermarking for identifying large language model outputs},
  journal   = {Nature},
  volume    = {634},
  pages     = {818--823},
  year      = {2024},
  month     = oct,
  doi       = {10.1038/s41586-024-08025-4},
  url       = {https://doi.org/10.1038/s41586-024-08025-4},
  publisher = {Springer Science and Business Media LLC}
}

@inproceedings{liu2024adaptive,
  author    = {Liu, Yepeng and Bu, Yuheng},
  title     = {Adaptive text watermark for large language models},
  year      = {2024},
  publisher = {JMLR.org},
  booktitle = {the 41st International Conference on Machine Learning},
  articleno = {1238},
  numpages  = {20},
  location  = {Vienna, Austria},
  series    = {ICML'24}
}

@article{liu2024survey,
  author     = {Liu, Aiwei and Pan, Leyi and Lu, Yijian and Li, Jingjing and Hu, Xuming and Zhang, Xi and Wen, Lijie and King, Irwin and Xiong, Hui and Yu, Philip},
  title      = {A Survey of Text Watermarking in the Era of Large Language Models},
  year       = {2024},
  issue_date = {February 2025},
  publisher  = {Association for Computing Machinery},
  address    = {New York, NY, USA},
  volume     = {57},
  number     = {2},
  issn       = {0360-0300},
  url        = {https://doi.org/10.1145/3691626},
  doi        = {10.1145/3691626},
  journal    = {ACM Comput. Surv.},
  month      = nov,
  articleno  = {47},
  numpages   = {36}
}

@inproceedings{zhao2024provable,
  title     = {Provable Robust Watermarking for {AI}-Generated Text},
  author    = {Xuandong Zhao and Prabhanjan Vijendra Ananth and Lei Li and Yu-Xiang Wang},
  booktitle = {The Twelfth International Conference on Learning Representations},
  year      = {2024},
  url       = {https://openreview.net/forum?id=SsmT8aO45L}
}

@inproceedings{zhao2025sok,
  author    = {Zhao, Xuandong and Gunn, Sam and Christ, Miranda and Fairoze, Jaiden and Fabrega, Andres and Carlini, Nicholas and Garg, Sanjam and Hong, Sanghyun and Nasr, Milad and Tramer, Florian and Jha, Somesh and Li, Lei and Wang, Yu-Xiang and Song, Dawn},
  booktitle = {2025 IEEE Symposium on Security and Privacy (SP)},
  title     = {{SoK}: Watermarking for {AI}-Generated Content},
  year      = {2025},
  volume    = {},
  issn      = {},
  pages     = {2621--2639},
  doi       = {10.1109/SP61157.2025.00178},
  url       = {https://doi.ieeecomputersociety.org/10.1109/SP61157.2025.00178},
  publisher = {IEEE Computer Society},
  address   = {Los Alamitos, CA, USA},
  month     = May
}

@article{davidson2018privacy,
  title   = {Privacy Pass: Bypassing Internet Challenges Anonymously},
  author  = {Alex Davidson and Ian Goldberg and Nick Sullivan and George Tankersley and Filippo Valsorda},
  journal = {Proceedings on Privacy Enhancing Technologies},
  year    = {2018},
  volume  = {2018},
  pages   = {164 - 180},
  doi     = {10.1515/POPETS-2018-0026}
}

@inproceedings{hou2024semstamp,
  title     = {{S}em{S}tamp: A Semantic Watermark with Paraphrastic Robustness for Text Generation},
  author    = {Hou, Abe  and
               Zhang, Jingyu  and
               He, Tianxing  and
               Wang, Yichen  and
               Chuang, Yung-Sung  and
               Wang, Hongwei  and
               Shen, Lingfeng  and
               Van Durme, Benjamin  and
               Khashabi, Daniel  and
               Tsvetkov, Yulia},
  booktitle = {the 2024 Conference of the North American Chapter of the Association for Computational Linguistics: Human Language Technologies (Volume 1: Long Papers)},
  month     = jun,
  year      = {2024},
  address   = {Mexico City, Mexico},
  publisher = {Association for Computational Linguistics},
  url       = {https://aclanthology.org/2024.naacl-long.226/},
  doi       = {10.18653/v1/2024.naacl-long.226},
  pages     = {4067--4082}
}

@inproceedings{liu2024unforgeable,
  title     = {An Unforgeable Publicly Verifiable Watermark for Large Language Models},
  author    = {Aiwei Liu and Leyi Pan and Xuming Hu and Shuang Li and Lijie Wen and Irwin King and Philip S. Yu},
  booktitle = {The Twelfth International Conference on Learning Representations},
  year      = {2024},
  url       = {https://openreview.net/forum?id=gMLQwKDY3N}
}

@inproceedings{duan2026pvmark,
  title     = {{PVMark}: Enabling Public Verifiability for {LLM} Watermarking Schemes},
  author    = {Haohua Duan and Liyao Xiang and Xin Zhang and Baochun Li and Bo Li},
  booktitle = {35th USENIX Security Symposium (USENIX Security 26)},
  year      = {2026},
  month     = aug,
  address   = {Baltimore, MD},
  publisher = {USENIX Association},
  pages     = {2007--2026},
  url       = {https://www.usenix.org/conference/usenixsecurity26/presentation/duan}
}

@inproceedings{hou2024ksemstamp,
  title     = {k-{S}em{S}tamp: A Clustering-Based Semantic Watermark for Detection of Machine-Generated Text},
  author    = {Hou, Abe  and
               Zhang, Jingyu  and
               Wang, Yichen  and
               Khashabi, Daniel  and
               He, Tianxing},
  booktitle = {Findings of the Association for Computational Linguistics: ACL 2024},
  month     = aug,
  year      = {2024},
  address   = {Bangkok, Thailand},
  publisher = {Association for Computational Linguistics},
  url       = {https://aclanthology.org/2024.findings-acl.98/},
  doi       = {10.18653/v1/2024.findings-acl.98},
  pages     = {1706--1715}
}

@inproceedings{wu2024resilient,
  author    = {Wu, Yihan and Hu, Zhengmian and Guo, Junfeng and Zhang, Hongyang and Huang, Heng},
  title     = {A resilient and accessible distribution-preserving watermark for large language models},
  year      = {2024},
  publisher = {JMLR.org},
  booktitle = {the 41st International Conference on Machine Learning},
  articleno = {2190},
  numpages  = {28},
  location  = {Vienna, Austria},
  series    = {ICML'24}
}

@inproceedings{zhu2024duwak,
  title     = {Duwak: Dual Watermarks in Large Language Models},
  author    = {Zhu, Chaoyi  and
               Galjaard, Jeroen  and
               Chen, Pin-Yu  and
               Chen, Lydia},
  booktitle = {Findings of the Association for Computational Linguistics: ACL 2024},
  month     = aug,
  year      = {2024},
  address   = {Bangkok, Thailand},
  publisher = {Association for Computational Linguistics},
  url       = {https://aclanthology.org/2024.findings-acl.678/},
  doi       = {10.18653/v1/2024.findings-acl.678},
  pages     = {11416--11436}
}

@inproceedings{guo2023domain,
  author    = {Guo, Junfeng and Li, Yiming and Wang, Lixu and Xia, Shu-Tao and Huang, Heng and Liu, Cong and Li, Bo},
  title     = {Domain watermark: effective and harmless dataset copyright protection is closed at hand},
  year      = {2023},
  publisher = {Curran Associates Inc.},
  address   = {Red Hook, NY, USA},
  booktitle = {the 37th International Conference on Neural Information Processing Systems},
  articleno = {2371},
  numpages  = {30},
  location  = {New Orleans, LA, USA},
  series    = {NIPS '23}
}

@inproceedings{cui2025robust,
  title     = {Robust Data Watermarking in Language Models by Injecting Fictitious Knowledge},
  author    = {Cui, Xinyue  and
               Wei, Johnny  and
               Swayamdipta, Swabha  and
               Jia, Robin},
  booktitle = {Findings of the Association for Computational Linguistics: ACL 2025},
  month     = jul,
  year      = {2025},
  address   = {Vienna, Austria},
  publisher = {Association for Computational Linguistics},
  url       = {https://aclanthology.org/2025.findings-acl.736/},
  doi       = {10.18653/v1/2025.findings-acl.736},
  pages     = {14292--14306},
  isbn      = {979-8-89176-256-5}
}

@inproceedings{zhao2023protecting,
  author    = {Zhao, Xuandong and Wang, Yu-Xiang and Li, Lei},
  title     = {Protecting language generation models via invisible watermarking},
  year      = {2023},
  publisher = {JMLR.org},
  booktitle = {the 40th International Conference on Machine Learning},
  articleno = {1774},
  numpages  = {13},
  location  = {Honolulu, Hawaii, USA},
  series    = {ICML'23}
}

@misc{llama2,
  title         = {Llama 2: Open Foundation and Fine-Tuned Chat Models},
  author        = {Meta},
  year          = {2023},
  eprint        = {2307.09288},
  archiveprefix = {arXiv},
  primaryclass  = {cs.CL},
  url           = {https://arxiv.org/abs/2307.09288},
  howpublished  = {arXiv:2307.09288}
}

@misc{gpt4,
  title         = {{GPT}-4 Technical Report},
  author        = {OpenAI},
  year          = {2024},
  eprint        = {2303.08774},
  archiveprefix = {arXiv},
  primaryclass  = {cs.CL},
  url           = {https://arxiv.org/abs/2303.08774},
  howpublished  = {arXiv:2303.08774}
}

@inproceedings{casacuberta2022sok,
  author    = {Casacuberta, Silvia and Hesse, Julia and Lehmann, Anja},
  booktitle = {2022 IEEE 7th European Symposium on Security and Privacy (EuroS\&P)},
  title     = {{SoK}: Oblivious Pseudorandom Functions},
  year      = {2022},
  volume    = {},
  issn      = {},
  pages     = {625-646},
  doi       = {10.1109/EuroSP53844.2022.00045},
  url       = {https://doi.ieeecomputersociety.org/10.1109/EuroSP53844.2022.00045},
  publisher = {IEEE Computer Society},
  address   = {Los Alamitos, CA, USA},
  month     = Jun
}

@misc{sun2024exploring,
  title         = {Exploring the Deceptive Power of {LLM}-Generated Fake News: A Study of Real-World Detection Challenges},
  author        = {Yanshen Sun and Jianfeng He and Limeng Cui and Shuo Lei and Chang-Tien Lu},
  year          = {2024},
  eprint        = {2403.18249},
  archiveprefix = {arXiv},
  primaryclass  = {cs.CL},
  url           = {https://arxiv.org/abs/2403.18249}
}

@inproceedings{lin2024malla,
  author    = {Zilong Lin and Jian Cui and Xiaojing Liao and XiaoFeng Wang},
  title     = {Malla: Demystifying Real-world Large Language Model Integrated Malicious Services},
  booktitle = {33rd USENIX Security Symposium (USENIX Security 24)},
  year      = {2024},
  isbn      = {978-1-939133-44-1},
  address   = {Philadelphia, PA},
  pages     = {4693--4710},
  url       = {https://www.usenix.org/conference/usenixsecurity24/presentation/lin-zilong},
  publisher = {USENIX Association},
  month     = aug
}

@article{raffel2020exploring,
  author     = {Raffel, Colin and Shazeer, Noam and Roberts, Adam and Lee, Katherine and Narang, Sharan and Matena, Michael and Zhou, Yanqi and Li, Wei and Liu, Peter J.},
  title      = {Exploring the limits of transfer learning with a unified text-to-text transformer},
  year       = {2020},
  issue_date = {January 2020},
  publisher  = {JMLR.org},
  volume     = {21},
  number     = {1},
  issn       = {1532-4435},
  journal    = {J. Mach. Learn. Res.},
  month      = jan,
  articleno  = {140},
  numpages   = {67}
}

@inproceedings{fan2019eli5,
  title     = {{ELI}5: Long Form Question Answering},
  author    = {Fan, Angela  and
               Jernite, Yacine  and
               Perez, Ethan  and
               Grangier, David  and
               Weston, Jason  and
               Auli, Michael},
  booktitle = {the 57th Annual Meeting of the Association for Computational Linguistics},
  month     = jul,
  year      = {2019},
  address   = {Florence, Italy},
  publisher = {Association for Computational Linguistics},
  url       = {https://aclanthology.org/P19-1346/},
  doi       = {10.18653/v1/P19-1346},
  pages     = {3558--3567}
}

@misc{cobbe2021training,
  title         = {Training Verifiers to Solve Math Word Problems},
  author        = {Karl Cobbe and Vineet Kosaraju and Mohammad Bavarian and Mark Chen and Heewoo Jun and Lukasz Kaiser and Matthias Plappert and Jerry Tworek and Jacob Hilton and Reiichiro Nakano and Christopher Hesse and John Schulman},
  year          = {2021},
  eprint        = {2110.14168},
  archiveprefix = {arXiv},
  primaryclass  = {cs.LG},
  url           = {https://arxiv.org/abs/2110.14168}
}

@misc{sanh2019distilbert,
  title         = {{DistilBERT}, a distilled version of {BERT}: smaller, faster, cheaper and lighter},
  author        = {Victor Sanh and Lysandre Debut and Julien Chaumond and Thomas Wolf},
  year          = {2020},
  eprint        = {1910.01108},
  archiveprefix = {arXiv},
  primaryclass  = {cs.CL},
  url           = {https://arxiv.org/abs/1910.01108}
}

@misc{qwen25,
  title         = {Qwen2.5 Technical Report},
  author        = {Qwen},
  year          = {2025},
  eprint        = {2412.15115},
  archiveprefix = {arXiv},
  primaryclass  = {cs.CL},
  url           = {https://arxiv.org/abs/2412.15115},
  howpublished  = {arXiv:2412.15115}
}

@inproceedings{jarecki2014roundoptimal,
  author    = {Jarecki, Stanislaw
               and Kiayias, Aggelos
               and Krawczyk, Hugo},
  title     = {Round-Optimal Password-Protected Secret Sharing and {T-PAKE} in the Password-Only Model},
  booktitle = {Advances in Cryptology -- ASIACRYPT 2014},
  year      = {2014},
  publisher = {Springer Berlin Heidelberg},
  address   = {Berlin, Heidelberg},
  pages     = {233--253},
  isbn      = {978-3-662-45608-8}
}

@article{shumailov2024collapse,
  author  = {Ilia Shumailov and Zakhar Shumaylov and Yiren Zhao and Nicolas Papernot and Ross J. Anderson and Yarin Gal},
  title   = {{AI} models collapse when trained on recursively generated data},
  year    = {2024},
  month   = {July},
  cdate   = {1719792000000},
  journal = {Nature},
  volume  = {631},
  number  = {8022},
  pages   = {755-759},
  url     = {https://doi.org/10.1038/s41586-024-07566-y}
}

@inproceedings{cai2025are,
  title     = {Are You Getting What You Pay For? {Auditing} Model Substitution in {LLM} {API}s},
  author    = {Will Cai and Tianneng Shi and Xuandong Zhao and Dawn Song},
  booktitle = {NeurIPS 2025 Workshop on Regulatable ML},
  year      = {2025},
  url       = {https://openreview.net/forum?id=thhrtv9P0s}
}

@inproceedings{gu2025auditing,
  title     = {Auditing Prompt Caching in Language Model {API}s},
  author    = {Chenchen Gu and Xiang Lisa Li and Rohith Kuditipudi and Percy Liang and Tatsunori Hashimoto},
  booktitle = {Forty-second International Conference on Machine Learning},
  year      = {2025},
  url       = {https://openreview.net/forum?id=gUj2fxQcLZ}
}

@misc{zhu2025auditing,
  title         = {Auditing Black-Box {LLM} {API}s with a Rank-Based Uniformity Test},
  author        = {Xiaoyuan Zhu and Yaowen Ye and Tianyi Qiu and Hanlin Zhu and Sijun Tan and Ajraf Mannan and Jonathan Michala and Raluca Ada Popa and Willie Neiswanger},
  year          = {2025},
  eprint        = {2506.06975},
  archiveprefix = {arXiv},
  primaryclass  = {cs.CR},
  url           = {https://arxiv.org/abs/2506.06975}
}

@inproceedings{wu2025synthetic,
  author    = {Wu, Yixin and Yang, Ziqing and Shen, Yun and Backes, Michael and Zhang, Yang},
  title     = {Synthetic artifact auditing: tracing {LLM}-generated synthetic data usage in downstream applications},
  year      = {2025},
  isbn      = {978-1-939133-52-6},
  publisher = {USENIX Association},
  address   = {USA},
  booktitle = {the 34th USENIX Conference on Security Symposium},
  articleno = {88},
  numpages  = {20},
  location  = {Seattle, WA, USA},
  series    = {SEC '25}
}

@inproceedings{amirizaniani2024auditllm,
  author    = {Amirizaniani, Maryam and Martin, Elias and Roosta, Tanya and Chadha, Aman and Shah, Chirag},
  title     = {{AuditLLM}: A Tool for Auditing Large Language Models Using Multiprobe Approach},
  booktitle = {the 33rd ACM International Conference on Information and Knowledge Management},
  year      = {2024},
  isbn      = {9798400704369},
  publisher = {Association for Computing Machinery},
  address   = {New York, NY, USA},
  doi       = {10.1145/3627673.3679222},
  pages     = {5174--5179},
  numpages  = {6},
  location  = {Boise, ID, USA},
  series    = {CIKM '24}
}

@inproceedings{chaum1992wallet,
  author    = {David Chaum and Torben P. Pedersen},
  title     = {Wallet Databases with Observers},
  year      = {1992},
  booktitle = {the 12th Annual International Cryptology Conference on Advances in Cryptology},
  publisher = {Springer},
  pages     = {89--105},
  isbn      = {3540573402},
  numpages  = {17}
}

@inproceedings{kissner2005privacy,
  author    = {Lea Kissner and Dawn Xiaodong Song},
  title     = {Privacy-Preserving Set Operations},
  year      = {2005},
  booktitle = {Advances in Cryptology -- CRYPTO 2005},
  publisher = {Springer},
  volume    = {3621},
  pages     = {241--257},
  doi       = {10.1007/11535218\_15}
}

@misc{chen2021codex,
  author        = {Mark Chen and Jerry Tworek and Heewoo Jun and Qiming Yuan and Henrique Ponde de Oliveira Pinto and Jared Kaplan and Harri Edwards and Yuri Burda and Nicholas Joseph and Greg Brockman and Alex Ray and Raul Puri and Gretchen Krueger and Michael Petrov and Heidy Khlaaf and Girish Sastry and Pamela Mishkin and Brooke Chan and Scott Gray and Nick Ryder and Mikhail Pavlov and Alethea Power and Lukasz Kaiser and Mohammad Bavarian and Clemens Winter and Philippe Tillet and Felipe Petroski Such and Dave Cummings and Matthias Plappert and Fotios Chantzis and Elizabeth Barnes and Ariel Herbert-Voss and William Hebgen Guss and Alex Nichol and Alex Paino and Nikolas Tezak and Jie Tang and Igor Babuschkin and Suchir Balaji and Shantanu Jain and William Saunders and Christopher Hesse and Andrew N. Carr and Jan Leike and Josh Achiam and Vedant Misra and Evan Morikawa and Alec Radford and Matthew Knight and Miles Brundage and Mira Murati and Katie Mayer and Peter Welinder and Bob McGrew and Dario Amodei and Sam McCandlish and Ilya Sutskever and Wojciech Zaremba},
  title         = {Evaluating Large Language Models Trained on Code},
  year          = {2021},
  url           = {https://arxiv.org/abs/2107.03374},
  eprint        = {2107.03374},
  archiveprefix = {arXiv},
  primaryclass  = {cs.LG}
}

@misc{kohli2025synthiddetector,
  author       = {Pushmeet Kohli},
  title        = {{SynthID} Detector --- a new portal to help identify {AI}-generated content},
  howpublished = {Google Blog},
  month        = {May 20},
  year         = {2025},
  url          = {https://blog.google/innovation-and-ai/products/google-synthid-ai-content-detector/}
}

@misc{iso19011,
  type        = {Standard},
  key         = {ISO 19011:2026},
  author      = {{International Organization for Standardization}},
  title       = {{ISO} 19011:2026 Guidelines for auditing management systems},
  year        = {2026},
  address     = {Geneva, Switzerland},
  institution = {International Organization for Standardization},
  url         = {https://www.iso.org/standard/19011}
}

@inproceedings{pang2024nofreelunch,
  author    = {Pang, Qi and Hu, Shengyuan and Zheng, Wenting and Smith, Virginia},
  title     = {No Free Lunch in {LLM} Watermarking: Trade-offs in Watermarking Design Choices},
  year      = {2024},
  isbn      = {9798331314385},
  publisher = {Curran Associates Inc.},
  address   = {Red Hook, NY, USA},
  booktitle = {Proceedings of the 38th International Conference on Neural Information Processing Systems},
  articleno = {4402},
  numpages  = {33},
  location  = {Vancouver, BC, Canada}
}

@inproceedings{linnartz1998sensitivity,
  author    = {Linnartz, Jean-Paul M. G. and van Dijk, Marten},
  title     = {Analysis of the Sensitivity Attack against Electronic Watermarks in Images},
  booktitle = {Information Hiding: Second International Workshop, IH 1998},
  editor    = {Aucsmith, David},
  series    = {Lecture Notes in Computer Science},
  volume    = {1525},
  pages     = {258--272},
  publisher = {Springer},
  year      = {1998},
  doi       = {10.1007/3-540-49380-8_18}
}

@inproceedings{kalker1998detector,
  author    = {Kalker, Ton and Linnartz, Jean-Paul M. G. and van Dijk, Marten},
  title     = {Watermark Estimation through Detector Analysis},
  booktitle = {Proceedings of the 1998 International Conference on Image Processing (ICIP 98)},
  volume    = {1},
  pages     = {425--429},
  publisher = {IEEE},
  year      = {1998},
  doi       = {10.1109/ICIP.1998.723516},
  isbn      = {0-8186-8821-1}
}

@inproceedings{adelsbach2001zk,
  author    = {Adelsbach, Andr{\'e} and Sadeghi, Ahmad-Reza},
  title     = {Zero-Knowledge Watermark Detection and Proof of Ownership},
  booktitle = {Information Hiding: Fourth International Workshop, IH 2001},
  editor    = {Moskowitz, Ira S.},
  series    = {Lecture Notes in Computer Science},
  volume    = {2137},
  pages     = {273--288},
  publisher = {Springer},
  year      = {2001},
  doi       = {10.1007/3-540-45496-9_20}
}

@article{adelsbach2003disclosure,
  author  = {Adelsbach, Andr{\'e} and Katzenbeisser, Stefan and Sadeghi, Ahmad-Reza},
  title   = {Watermark Detection with Zero-Knowledge Disclosure},
  journal = {Multimedia Systems},
  volume  = {9},
  number  = {3},
  pages   = {266--278},
  year    = {2003},
  doi     = {10.1007/s00530-003-0098-z}
}

@inproceedings{troncoso2006sensitivity,
  author    = {Troncoso-Pastoriza, Juan Ram{\'o}n and P{\'e}rez-Gonz{\'a}lez, Fernando},
  title     = {Zero-Knowledge Watermark Detector Robust to Sensitivity Attacks},
  booktitle = {Proceedings of the 8th ACM Workshop on Multimedia and Security},
  pages     = {97--107},
  publisher = {ACM},
  year      = {2006},
  doi       = {10.1145/1161366.1161383},
  isbn      = {1-59593-493-6}
}

@inproceedings{barni2014smart,
  author    = {Barni, Mauro and Comesa{\~n}a-Alfaro, Pedro and P{\'e}rez-Gonz{\'a}lez, Fernando and Tondi, Benedetta},
  title     = {Are You Threatening Me? Towards Smart Detectors in Watermarking},
  booktitle = {Media Watermarking, Security, and Forensics 2014},
  series    = {Proceedings of SPIE},
  volume    = {9028},
  pages     = {51--62},
  articleno = {902806},
  publisher = {SPIE},
  year      = {2014},
  doi       = {10.1117/12.2036415}
}

@misc{eu_2024_1689_art50,
  title  = {{Regulation ({EU}) 2024/1689 Laying Down Harmonised Rules on Artificial Intelligence, Article 50: Transparency Obligations}},
  author = {{European Parliament and Council}},
  year   = {2024},
  url    = {https://eur-lex.europa.eu/eli/reg/2024/1689/oj/eng},
  note   = {Article 50 applicable from 2 August 2026}
}

@misc{ca_sb942,
  author       = {{California State Legislature}},
  title        = {{SB} 942, California {AI} Transparency Act},
  year         = {2024},
  howpublished = {Senate Bill No. 942},
  note         = {Chapter 291, Statutes of 2024; amended by AB 853 (2025), operative 2 August 2026},
  url          = {https://leginfo.legislature.ca.gov/faces/billNavClient.xhtml?bill_id=202320240SB942}
}

@misc{c2pa,
  author       = {{Coalition for Content Provenance and Authenticity (C2PA)}},
  title        = {{C2PA Technical Specification}},
  howpublished = {\url{https://spec.c2pa.org/}},
  year         = {2026},
  note         = {Version 2.4, Accessed July 29, 2026}
}

@inproceedings{jovanovic2024stealing,
  title     = {Watermark Stealing in Large Language Models},
  author    = {Jovanovi{\'c}, Nikola and Staab, Robin and Vechev, Martin},
  booktitle = {Proceedings of the 41st International Conference on Machine Learning},
  pages     = {22956--22978},
  year      = {2024},
  editor    = {Salakhutdinov, R. and Kolter, Z. and Hardt, M. and Etzioni, O. and Lee, J. and Habibullah, N. and Imkel, G.},
  volume    = {235},
  series    = {Proceedings of Machine Learning Research},
  month     = jul,
  publisher = {PMLR},
  url       = {https://proceedings.mlr.press/v235/jovanovic24a.html}
}

\appendix
\section{Proof of Correctness for Unbiased Sampling}
\label{app:unbiased-sampling}

We prove that the rejection sampling procedure described in the main paper samples exactly from the target watermarked distribution $q(t)$, where $q(t)\propto p(t)\cdot \exp(\delta)^{\indicator(t\in G)}$ and $\indicator(\cdot)$ is the indicator function.

\begin{proof}
  Let $R = \mathcal{V} \setminus G$ be the set of red tokens.
  The normalized target distribution is
  \begin{equation}
    q(t) = \frac{1}{Z} \cdot p(t) \cdot \exp(\delta)^{\indicator(t\in G)},
  \end{equation}
  where the normalizing constant is
  \begin{equation}
    \label{eq:appendix-partition-function}
    Z = \sum_{t'\in R} p(t') + \exp(\delta) \sum_{t'\in G} p(t').
  \end{equation}
  In one proposal step, the algorithm draws $t$ from $p(t)$ and accepts it with probability $1$ for green tokens and $\exp(-\delta)$ for red tokens.
  Thus, the probability that one proposal selects and accepts token $t$ is
  \begin{align*}
    \Pr(\text{propose }t\text{ and accept})
    &=
    \begin{cases}
      p(t),             & t \in G, \\
      p(t) \exp(-\delta), & t \in R.
    \end{cases}
  \end{align*}
  The total acceptance probability in one proposal step is therefore
  \begin{align*}
    \Pr(\text{accept}) &= \sum_{t \in G} p(t) + \sum_{t \in R} p(t) \exp(-\delta) \\
    &= \exp(-\delta) \left( \exp(\delta) \sum_{t \in G} p(t) + \sum_{t \in R} p(t) \right).
  \end{align*}
  By Equation~\ref{eq:appendix-partition-function}, $\Pr(\text{accept}) = Z\exp(-\delta)$.
  Because proposals are repeated independently until one is accepted, the final output distribution equals the distribution of one proposal conditioned on acceptance:
  \begin{equation}
    \Pr(\text{output}=t) = \frac{\Pr(\text{propose }t\text{ and accept})}{\Pr(\text{accept})}.
  \end{equation}
  Substituting the expressions above gives
  \begin{align*}
    \Pr(\text{output}=t)
    &=
    \begin{cases}
      p(t)\exp(\delta)/Z, & t \in G, \\
      p(t)/Z,             & t \in R,
    \end{cases} \\
    &= q(t).
  \end{align*}
  Hence the rejection sampling process samples exactly from the desired target distribution.
\end{proof}

\section{Security Proof}
\label{app:security-proof}

We prove the three properties of Section~\ref{sec:threat-model} for the 2HashDH instantiation~\cite{jarecki2014roundoptimal,rfc9497}.
Let $\mathbb{G}$ be a group of prime order $q$ with generator $g$, let $H_1:\{0,1\}^{*}\to\mathbb{G}$ and $H_2$ be hash functions modeled as random oracles, and let $\mathsf{sk}=s\in\mathbb{Z}_q^{*}$ with $\mathsf{pk}=g^{s}$.
The scheme computes $\mathcal{F}(s,x)=H_2\!\left(x,H_1(x)^{s}\right)$, with $\mathsf{Blind}(x)$ drawing $r\leftarrow\mathbb{Z}_q^{*}$ and returning $\tilde{x}=H_1(x)^{r}$.
On $\tilde{\mathbf{x}}=(\tilde{x}_1,\ldots,\tilde{x}_n)$, $\mathsf{BlindEvaluate}(s,\tilde{\mathbf{x}})$ returns $\tilde{\mathbf{y}}=(\tilde{x}_1^s,\ldots,\tilde{x}_n^s)$ and a batched DLEQ proof $\pi$; after verification, $\mathsf{Finalize}(x_i,\tilde{y}_i,r_i)=H_2\!\left(x_i,\tilde{y}_i^{1/r_i}\right)$.
For a fixed valid $\mathsf{SR}$, let $\mathsf{View}_{\mathcal{D}^{*}}(T)$ denote $\mathcal{D}^{*}$'s cryptographic view of an execution with an honest auditor holding $T$.

\subsection{Input Privacy}

\begin{theorem}
  \label{thm:privacy}
  For every detection service $\mathcal{D}^{*}$ and every pair of texts $T_0,T_1$ inducing the same number $n$ of inputs, the statistical distance between $\mathsf{View}_{\mathcal{D}^{*}}(T_0)$ and $\mathsf{View}_{\mathcal{D}^{*}}(T_1)$ is at most $2n/q$ and is therefore negligible.
\end{theorem}

\begin{proof}
  The messages $\mathcal{D}^{*}$ receives are $\tilde{x}_1,\ldots,\tilde{x}_n$ with $\tilde{x}_i=H_1(x_i)^{r_i}$ and $r_i$ drawn independently and uniformly from $\mathbb{Z}_q^{*}$.
  Since $q$ is prime, every element of $\mathbb{G}\setminus\{1\}$ generates $\mathbb{G}$, so for any fixed $h\in\mathbb{G}\setminus\{1\}$ the map $r\mapsto h^{r}$ is a bijection from $\mathbb{Z}_q^{*}$ onto $\mathbb{G}\setminus\{1\}$.
  Conditioned on $H_1(x_i)\ne 1$ for every input, each $\tilde{x}_i$ is therefore uniform on $\mathbb{G}\setminus\{1\}$ and independent of $x_i$, and the tuple is uniform on $(\mathbb{G}\setminus\{1\})^{n}$ for either text.
  Everything else in the view is computed by $\mathcal{D}^{*}$ from these messages and its own randomness.
  The RFC~9497 blinding procedure rejects when $H_1(x_i)=1$~\cite{rfc9497}; in the random-oracle model, a union bound places the probability of such an event in each execution at most $n/q$.
  The statistical distance between the two views is therefore at most $2n/q$, which is negligible in the security parameter.
\end{proof}

\subsection{Detection Soundness}

\begin{theorem}
  \label{thm:soundness}
  Assume the batched DLEQ proof system used by the VOPRF is computationally sound.
  For every valid $\mathsf{SR}$, text $T$, and PPT detection service $\mathcal{D}^{*}$, the probability that an honest auditor completes without aborting and outputs a result different from $\mathsf{Result}_{\mathsf{SR}}(T)$ is at most the \mbox{negligible function $\varepsilon_{\mathrm{batch}}(\lambda)$}.
\end{theorem}

\begin{proof}
  Fix $\mathsf{SR}$, $T$, and $\mathcal{D}^{*}$, and let $\mathsf{pk}=g^s$ be the public key fixed by $\mathsf{SR}$.
  Suppose $\mathcal{A}$ completes without aborting, so $\mathsf{VerifyBatch}(\mathsf{pk},\allowbreak\tilde{\mathbf{x}},\allowbreak\tilde{\mathbf{y}},\allowbreak\pi)=\mathit{Accept}$.
  By soundness of the batched proof system, except with probability $\varepsilon_{\mathrm{batch}}(\lambda)$, acceptance implies $\tilde{y}_i=\tilde{x}_i^{s}$ for every $i$.
  Since the honest auditor formed $\tilde{x}_i=H_1(x_i)^{r_i}$, every finalized value then satisfies
  \begin{align*}
    \mathsf{Finalize}(x_i,\tilde{y}_i,r_i)
    &= H_2\!\left(x_i,\tilde{y}_i^{1/r_i}\right) \\
    &= H_2\!\left(x_i,H_1(x_i)^{s}\right) \\
    &= \mathcal{F}(s,x_i).
  \end{align*}
  The honest auditor derives every input prescribed for $T$ by $\mathsf{SR}$ and computes the green count, $p$-value, and decision deterministically from these finalized values under its detector specification.
  Thus accepting a result different from $\mathsf{Result}_{\mathsf{SR}}(T)$ requires a violation of batched-proof soundness and occurs with probability at most $\varepsilon_{\mathrm{batch}}(\lambda)$.
\end{proof}

With individual proofs instead of batching, the same argument applies per pair and a union bound gives $n\cdot\varepsilon_{\mathrm{DLEQ}}(\lambda)$.

\subsection{Transferable Evidence}

Recall that $\mathsf{Arbitrate}(T,\tau,\mathsf{SR})$ performs the four checks of Section~\ref{sec:analysis-evidence}.

\begin{theorem}
  \label{thm:evidence}
  Assuming unforgeability of the setup-record signature and soundness of the batched DLEQ proof system, $\mathsf{Arbitrate}$ satisfies Definition~\ref{def:evidence}.
\end{theorem}

\begin{proof}
  The four checks are polynomial-time and use only public algorithms and the supplied $T$, $\tau$, and $\mathsf{SR}$.
  Thus $\mathsf{Arbitrate}$ requires neither the watermarking key nor interaction with the provider.
  On the transcript of an honest audit, check~1 authenticates the setup record identified in $\tau$.
  The auditor derived every input prescribed for $T$, used valid blinding randomness, and recorded one entry for each input, so check~2 passes.
  Completeness of the batched DLEQ proof system makes check~3 pass, and VOPRF correctness makes check~4 reconstruct the same outputs, green count, $p$-value, and decision as the auditor.
  Hence $\mathsf{Arbitrate}$ returns $\mathsf{Result}_{\mathsf{SR}}(T)$ on every honestly completed audit.

  Now let a PPT adversary produce a transcript $\tau$ that $\mathsf{Arbitrate}$ accepts.
  By signature unforgeability, except with negligible probability, the setup record accepted in check~1 was signed by $\mathcal{O}$.
  Check~2 establishes a one-to-one correspondence between the complete input set derived from $T$ and the recorded entries.
  After indexing each entry by its matched input, the check also validates $r_i\in\mathbb{Z}_q^{*}$ and $\tilde{x}_i=H_1(x_i)^{r_i}$ for every $i$.
  By soundness of the batched DLEQ proof system, acceptance in check~3 implies, except with negligible probability, that $\tilde{y}_i=\tilde{x}_i^s$ for every $i$, where $s=\log_g\mathsf{pk}$.
  Therefore check~4 obtains
  \begin{equation*}
    y_i
    =H_2\!\left(x_i,\tilde{y}_i^{1/r_i}\right)
    =H_2\!\left(x_i,H_1(x_i)^s\right)
    =\mathcal{F}(s,x_i)
  \end{equation*}
  for every input and deterministically recomputes the statistic and decision prescribed by $\mathsf{SR}$.
  Conditioned on an authenticated $\mathsf{SR}$, an accepted arbitration can therefore return a result different from $\mathsf{Result}_{\mathsf{SR}}(T)$ only by violating batched-proof soundness.
  Together with signature unforgeability for check~1, this proves Definition~\ref{def:evidence}.
\end{proof}

\section{Analysis of the Fixed-\texorpdfstring{$k$}{k} Construction}
\label{app:forgery-cost}

This appendix derives the reference success probability and cost of the fixed-$k$ construction in Section~\ref{sec:analysis-forgery} under conditions (D1)--(D3).
Condition (D2) provides the adaptive color feedback used to select tokens, while (D3) lets the producer determine whether the completed text passes.
Condition (D1), together with the freshness refinement below, supplies the Bernoulli reference model for the success and cost calculations.
The construction can still be executed against an interface whose concrete color responses depart from (D1), but the formulas below then serve only as reference predictions.

\subsection{Freshness and Green Count}

The producer fixes the text from left to right, evaluating up to $k$ candidate continuations at each position and selecting the first green candidate, or the most probable candidate if all $k$ are red.
For the reference calculation, we refine candidate selection to enforce the freshness premise in (D1).
The producer maintains a set $H$ of all pairs it has previously queried under the selected $\mathsf{SR}$, including in earlier attempts.
At position $i$, it scans tokens in descending model probability and forms a list of the first $k$ candidates $t$ for which $(c_i,t)\notin H$.
If fewer than $k$ such candidates are available, the attempt aborts.
Each queried pair is added to $H$ before construction proceeds.
When a context recurs, this rule may move the candidate list beyond the original top $k$, but checking $H$ is local and adds no detector evaluations.
Thus, even when a context recurs, every queried pair is fresh and every selected pair is unique.
Let $n$ denote the number of selected pairs in the completed text.
Under (D1), the candidate colors are conditionally independent Bernoulli trials with parameter $\gamma$.

The probability $\theta(k)$ that a selected pair is green is
\begin{equation}
  \theta(k)=1-(1-\gamma)^{k},
\end{equation}
and the green count $Z$ among the $n$ selected pairs satisfies $Z\sim\mathrm{Binomial}(n,\theta(k))$.

The detector's critical green count is
\begin{equation}
  g^{*}=\min\left\{g\in\{0,\ldots,n+1\}:\Pr[\mathrm{Binomial}(n,\gamma)\ge g]\le\alpha\right\}
\end{equation}
where $g^{*}=n+1$ means that no attainable count passes.
Under (D3), the construction's reference-model success \mbox{probability is therefore}
\begin{equation}
  \Pr[Z\ge g^{*}]
  =\Pr[\mathrm{Binomial}(n,\theta(k))\ge g^{*}].
\end{equation}

\subsection{Pair Evaluations and Rounds}

Let $Q$ denote the total number of pair evaluations.
At position $i$, the construction reaches candidate $j$ exactly when its first $j-1$ candidates are red.
Thus, with $N_i$ denoting the number of evaluations at that position,
\begin{equation}
  \mathbb{E}[N_i]
  =\sum_{j=1}^{k}\Pr[N_i\ge j]
  =\sum_{j=0}^{k-1}(1-\gamma)^j
  =\frac{\theta(k)}{\gamma}.
\end{equation}
Summing over the $n$ selected pairs gives $\mathbb{E}[Q]/n=\theta(k)/\gamma$.

In the sequential realization used in our evaluation, each evaluation is sent only after the preceding answer is known, so the number $B$ of adaptive service rounds equals $Q$.
For fixed $k$, $n\le B\le kn$, and hence this construction uses $\Theta(n)$ rounds.
An honest audit instead submits all $n$ final pairs in one batch and evaluates each once.
The construction costs above do not include a transferable record for the completed text; obtaining one requires a subsequent ordinary audit batch under the transcript format evaluated in this paper.

\subsection{Interface Boundary}

The construction requires the requester to infer the exact color of each queried pair before choosing its next candidate.
Exact per-pair output provides this information directly, while some aggregate interfaces may reveal it through isolation or differences between queries.
Interfaces that prevent such inference do not satisfy (D2).

\section{Watermarking Parameter Selection}
\label{app:parameter-selection}

\begin{figure*}
  \centering
  \includegraphics{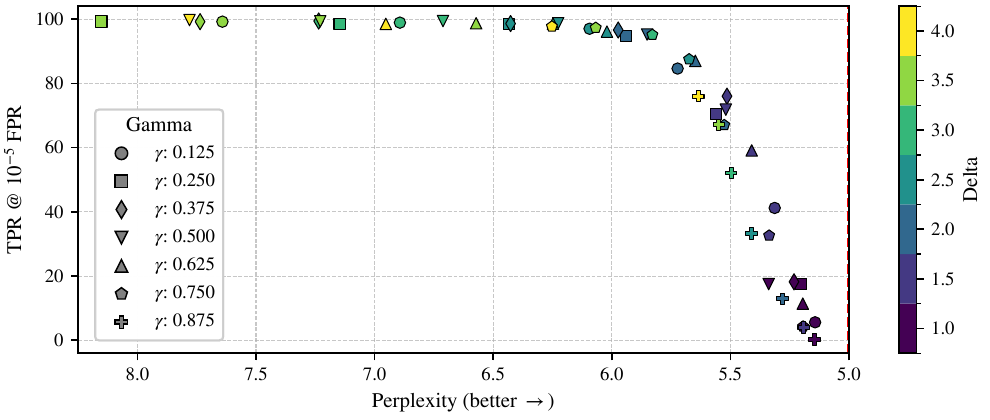}
  \includegraphics{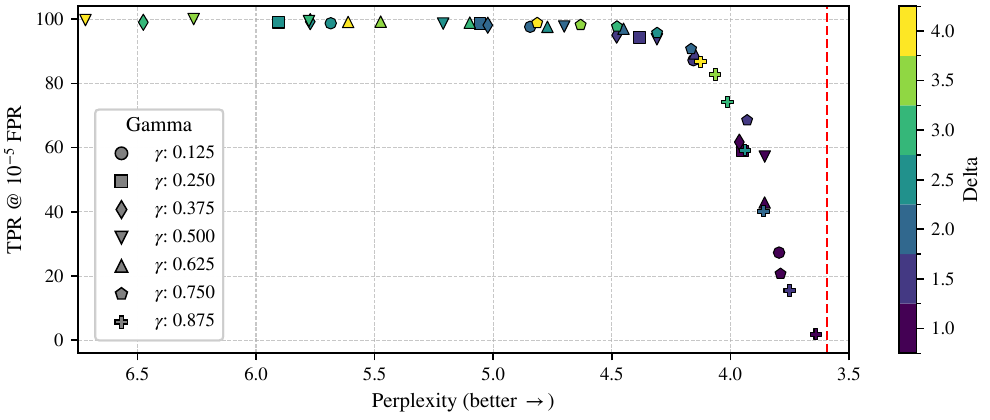}
  \caption{Trade-off between perplexity and true positive rate at the nominal significance level $\alpha=10^{-5}$ with default multinomial (top) and top-50 (bottom) sampling. The context length is set to $h=2$. Points outside the displayed perplexity range are omitted for visibility. The red dashed line indicates the perplexity of non-watermarked texts using the same sampling setting.}
  \label{fig:ppl-tpr}
\end{figure*}

\noindent\textbf{Experimental setup.}
We use a grid search to evaluate the trade-off between generation quality and watermark detectability across different watermark parameter settings.
Figure~\ref{fig:ppl-tpr} reports the sweep at $h=2$, matching the default context length used in the main text.
We employ two different sampling strategies: default multinomial sampling and top-50 multinomial sampling with nucleus sampling ($p=0.9$).
The latter setting is commonly used in text generation and gives a higher-quality operating point in our experiments.
We use 1000 prompts sampled and truncated from the C4 dataset for evaluation.
All texts are generated using the same Qwen2.5-3B model and each contains about 200 tokens.

\noindent\textbf{Results.}
The results are summarized in Figure~\ref{fig:ppl-tpr}, where the red dashed lines represent the perplexity of non-watermarked texts.
Increasing the watermark strength (i.e., using a larger $\delta$) generally improves TPR at the cost of higher perplexity.
As the green-token threshold $\gamma$ increases, a larger bias strength $\delta$ is needed to maintain high detectability.
When $\gamma$ is very small or very large, e.g., $\gamma=0.125$ or $\gamma=0.875$, the trade-off becomes more sensitive to $\delta$.
For $h=2$, the selected default is $\gamma=0.5$, $\delta=3.0$; reducing $\delta$ to 2.5 gives the quality-oriented alternative reported in Section~\ref{sec:evaluation-performance}.

\noindent\textbf{Differences from KGW settings.}
KGW's recommended parameters, for both the LeftHash and SelfHash variants, do not provide the best operating point for VOW in our evaluation, despite the schemes sharing a logit-biasing structure.
Their partition functions and sampling implementations differ, so we do not assume that the same values of $\gamma$ and $\delta$ transfer directly.
We instead select VOW's parameters empirically after fixing $h=2$, as described above.

\section{Data Preparation Details}
\label{app:data-preparation}

We use the news-like subset of the C4 dataset for generating watermarked text samples.
Specifically, we randomly sample 500 texts from this subset and use the first 60 tokens as the prompt for text generation.
We use a default multinomial sampling strategy with a temperature of 0.7.
The generated texts contain about 209 tokens on average and are used for evaluating detectability and text quality.

Watermarked text samples for robustness evaluation are generated using the ELI5 dataset~\cite{fan2019eli5}, which contains long-form answers to open-ended questions.
We randomly select 500 questions as prompts for data preparation, and use Qwen2.5-3B-Instruct with a top-50 sampling strategy and a temperature of 0.7 to generate watermarked text samples.
The prompt template is as follows:

\begin{center}
  \par
  \addvspace{1em}
  \fbox{\parbox{.42\textwidth}{Explain the following question like I'm 5 years old. Use very simple language, short sentences, and analogies a child can understand.

      Question: \{question\}
    }}
  \par
  \addvspace{1em}
\end{center}

The prompt used for paraphrasing is taken from prior work~\cite{kirchenbauer2024reliability}:

\begin{center}
  \par
  \addvspace{1em}
  \fbox{\parbox{.42\textwidth}{As an expert copy-editor, please rewrite the following text in your own voice while ensuring that the final output contains the same information as the original text and has roughly the same length. Please paraphrase all sentences and do not omit any crucial details. Additionally, please take care to provide any relevant information about public figures, organizations, or other entities mentioned in the text to avoid any potential misunderstandings or biases.

      Text: \{text\}
    }}
  \par
  \addvspace{1em}
\end{center}

\noindent\textbf{Attack fidelity.}
We measure semantic similarity between the original and attacked texts using the text-embedding-3-large model, and the ratio of attacked to original character length to quantify compression or expansion.
Table~\ref{tab:attack-similarity} shows substantial semantic similarity overall and approximate character-length preservation under paraphrasing.
PDW is the main exception: its longer source texts are compressed substantially by both paraphrasers and have lower semantic similarity.

\begin{table}[t]
  \centering
  \caption{Average semantic similarity and character-length retention between watermarked texts and their attacked versions. Retention below one \mbox{indicates compression}.}
  \label{tab:attack-similarity}
  \footnotesize
  \setlength{\tabcolsep}{4.5pt}
  \begin{tabular}{cccc}
    \toprule
    Method
    & \makecell[c]{Synonym\\Similarity}
    & \makecell[c]{GPT-3.5 Paraphrase\\Sim.\ (Length)}
    & \makecell[c]{GPT-5.6 Paraphrase\\Sim.\ (Length)} \\
    \midrule
    VOW ($h=1$) & 0.936 & 0.891 (1.135) & 0.900 (1.073) \\
    VOW ($h=2$) & 0.936 & 0.887 (1.091) & 0.890 (1.059) \\
    VOW ($h=4$) & 0.938 & 0.889 (1.091) & 0.889 (1.058) \\
    LeftHash    & 0.938 & 0.888 (1.093) & 0.891 (1.059) \\
    SelfHash    & 0.935 & 0.892 (1.126) & 0.896 (1.068) \\
    RDF         & 0.936 & 0.891 (1.108) & 0.895 (1.065) \\
    PDW         & 0.959 & 0.771 (0.377) & 0.735 (0.352) \\
    UPV         & 0.938 & 0.885 (1.086) & 0.889 (1.071) \\
    \bottomrule
  \end{tabular}
\end{table}

\section{Foundational Validation}
\label{app:foundational-validation}

Our statistical detector relies on the partition predicate having the intended marginal green rate and on the resulting binomial $p$-values being calibrated on non-watermarked text.
We evaluate these two conditions separately.

\begin{figure}[t]
  \centering
  \includegraphics{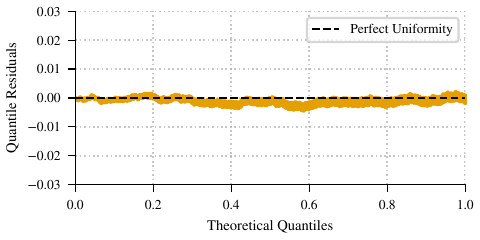}
  \caption{Marginal balance of the partitioning predicate $\mathrm{IsGreen}$. The $x$-axis shows theoretical uniform quantiles, and the $y$-axis shows the observed quantile residuals from \num{1000000} sampled contexts. The dashed line marks zero residual under perfect uniformity.}
  \label{fig:pseudorandomness}
\end{figure}

\noindent\textbf{Partition balance.}
Our detection process returns a $p$-value assuming that the event $t\in G$ occurs with probability $\gamma$ under the stated null model.
To test the predicate's marginal balance, we randomly sample \num{1000000} contexts from the C4 dataset and draw an independent random watermarking key for each vocabulary partition.
For each partition, we conduct a Pearson's chi-squared test to check whether the observed distribution of green tokens matches the expected distribution.
Each chi-squared test returns a $p$-value, and we expect the $p$-values to be uniformly distributed in $[0, 1]$.
The quantile residual plot in Figure~\ref{fig:pseudorandomness} shows the difference between the observed distribution and the expected uniform distribution.
The observed quantile residuals are centered around zero with low variance, indicating that the observed marginal distribution closely follows the expected uniform distribution for the sampled contexts and keys.
The following direct FPR experiment separately evaluates calibration of the complete detector on unwatermarked texts.

\begin{figure}[t]
  \centering
  \includegraphics{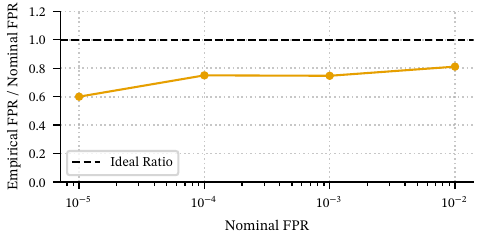}
  \caption{Ratio of empirical to nominal false-positive rates at the four reported operating points for VOW with $h=2$. The dashed line indicates the ideal ratio of 1.}
  \label{fig:fpr}
\end{figure}

\noindent\textbf{Detector reliability.}
As Fernandez et al.~\cite{fernandez2023three} note, empirical false positive rates (FPRs) of watermark detectors can be higher than theoretical ones due to pseudorandomness and repetitive texts.
We therefore conduct a large-scale calibration check for the watermark detector.
We compare the empirical and nominal FPRs of the watermark detector using \num{1000000} sampled texts of length 255 tokens from the C4 dataset.
Figure~\ref{fig:fpr} plots the empirical-to-nominal FPR ratio for the four $h=2$ operating points reported in Table~\ref{tab:fpr-calibration}.
The table also gives the direct false-positive counts for VOW and the other $\alpha$-indexed detectors.
At the strictest reported level, $\alpha=10^{-5}$, VOW produces 6 false positives in \num{1000000} samples, close to the nominal reference count of 10.
At the three larger levels, its empirical FPR remains below the nominal level, as permitted by the conservative discrete exact-binomial test.

\section{UPV Detection Operating Point and Cross-Network Agreement}
\label{app:upv-detection-reliability}

\begin{table*}[t]
  \centering
  \caption{Joint decisions of UPV's public detection network and a generation-side color test on \num{1000000} unwatermarked C4 texts.
    The color test counts tokens classified as green by the generation network and applies a $z$-score threshold of 4.0; the public detector uses its default threshold of 0.5.
  The two columns are not ground-truth labels.}
  \label{tab:upv-fpr}
  \footnotesize
  \begin{tabular}{lSS}
    \toprule
    & {Positive by color test} & {Negative by color test} \\
    \midrule
    Positive by public detector & 1437 & 311 \\
    Negative by public detector & 6483 & 991769 \\
    \bottomrule
  \end{tabular}
\end{table*}

UPV~\cite{liu2024unforgeable} uses two separate neural networks for vocabulary partitioning and watermark detection, respectively.
The generation network assigns a green/red status to each token in context, whereas the public detection network directly classifies a complete text.
We first measure the public detector's empirical FPR at its default threshold on \num{1000000} unwatermarked C4 texts.
It flags 1748 texts, giving an empirical FPR of $1.748\times 10^{-3}$.
Because this classifier threshold has no nominal $\alpha$ mapping, this is UPV's relevant operating-point FPR in our evaluation and is reported separately from Table~\ref{tab:fpr-calibration}.

Table~\ref{tab:upv-fpr} additionally compares the public detector with a text-level rule constructed from the generation network's token colors.
The color rule flags 7920 texts, of which 1437 are also flagged by the public detector; the two rules disagree on 6794 texts, or 0.6794\% of the sample.
This disagreement is not another FPR estimate and does not by itself identify erroneous predictions, because the two resulting rules depend on different learned networks and thresholds and were not calibrated to make identical decisions.
Instead, it shows that the public detector's output is not interchangeable with the generation-side color statistic; the $1.748\times 10^{-3}$ public-detector FPR, rather than the cross-network disagreement rate, is the quantity used when interpreting UPV's detection results.

\section{Contextualizing KGW Robustness Under Paraphrasing}
\label{app:kgw-paraphrasing}

Our robustness evaluation (Section~\ref{sec:evaluation-robustness}) yields lower KGW AUC under GPT-3.5 Turbo paraphrasing than the original KGW evaluation~\cite{kirchenbauer2024reliability}.
The prior work reported AUC above 0.85 at 200 tokens and above 0.9 at 600 tokens, whereas both KGW variants remain below 0.8 under our generation and attack settings.
We compare the experimental setups to contextualize this difference, not to assign it a cause.

Several parameters in our evaluation were kept consistent with the prior work to facilitate comparison.
The core KGW scheme settings were identical ($\gamma=0.25$ and $\delta=2.0$).
We used the same paraphrasing prompt and sampling setup (a temperature of 0.7), and the resulting text lengths were comparable (around 500--1000 tokens).
The base watermarking model differs (our Qwen2.5-3B-Instruct versus their Llama-7B), as does the prompt dataset (our ELI5 versus their C4).
Moreover, although both evaluations used a hosted system named ``GPT-3.5-Turbo'', changes to that system between the two evaluations could affect paraphrasing strength.
These are plausible methodological differences, but the available evidence does not identify the cause of the AUC gap.

\section{Impact of Watermarking Parameters on Generation Speed}
\label{app:parameter-impact-on-speed}

\begin{figure}[t]
  \centering
  \includegraphics{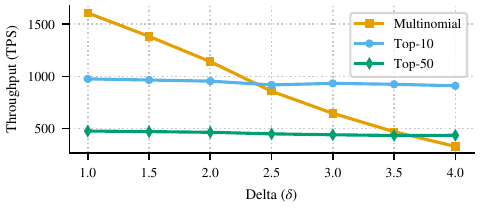}
  \includegraphics{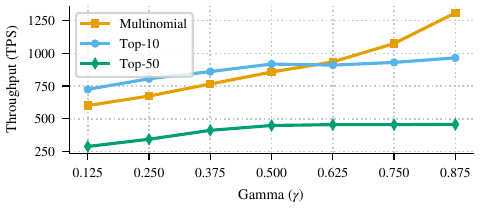}
  \caption{End-to-end throughput of VOW under different parameter settings. $\gamma$ is fixed to 0.5 in the top figure, and $\delta$ is fixed to 2.5 in the bottom figure.}
  \label{fig:watermark-tps}
\end{figure}

We report the end-to-end throughput under different sampling strategies in Figure~\ref{fig:watermark-tps}.
The experimental platform is the same as described in Section~\ref{sec:evaluation-overhead}.
Three sampling strategies are evaluated: default multinomial sampling and top-k sampling with $k=10$ and $k=50$.
As $\delta$ increases or the green-token threshold $\gamma$ decreases, the generation speed tends to decrease for all evaluated sampling strategies.
For top-k sampling, this is because a larger $\delta$ increases the number of candidate tokens that need to be checked after filtering.
For the default multinomial sampler without top-k filtering, let $\rho$ denote the probability mass assigned to green tokens in the original distribution.
The expected number of calls to the $\mathrm{IsGreen}$ predicate per generated token is:
\begin{equation}
  \frac{\exp(\delta)}{1 + (\exp(\delta) - 1) \rho}
  = \frac{1}{\rho + (1 - \rho) \exp(-\delta)}.
\end{equation}
Here, $\rho$ is approximately $\gamma$ in expectation over the random predicate.
Thus, a larger $\delta$ or a smaller $\rho$ leads to a higher expected number of calls, and consequently a lower throughput.

For top-k sampling, our two-stage optimization first forms $C_k$ from every token that could enter the post-bias top-$k$ set and then iterates over this candidate set with early stopping to construct the final token distribution.
To investigate the impact of the watermark parameter $\delta$ and sampling parameter $k$ on this optimization, we measure the average size of the candidate set $C_k$ and the number of tokens evaluated before early stopping.

The results are summarized in Figure~\ref{fig:candidate-set-size} in the main text.
As $k$ increases, both the average size of the candidate set $C_k$ and the average number of iterated tokens increase.
This is expected since a larger $k$ allows more tokens to be included in $C_k$.
However, the average number of iterated tokens grows much more slowly than the size of $C_k$, especially when $\delta$ is large.
The slopes of the curves for the candidate set size are 1.97, 3.77, 7.53, and 14.31 for $\delta=1, 2, 3, 4$, respectively, while the slopes for the average number of iterated tokens are within the range of 1.88 to 2.13.
Additionally, the number of evaluated tokens has lower variance than the candidate-set size.
This demonstrates that our early stopping strategy is effective in reducing the number of tokens that need to be processed, thereby improving generation speed.

\section{Communication Overhead of VOW}
\label{app:communication-overhead}

\begin{figure}[t]
  \centering
  \includegraphics{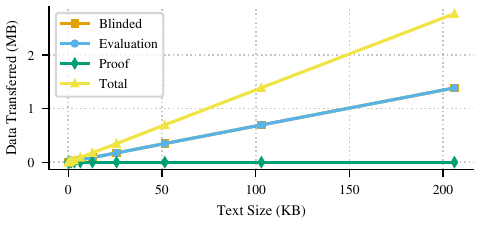}
  \caption{Per-audit VOW protocol payload with varying input text sizes, excluding the one-time setup record and transport framing.}
  \label{fig:communication-overhead}
\end{figure}

As shown in Figure~\ref{fig:communication-overhead}, the measured per-audit protocol payload scales linearly with the input text size.
It consists of the auditor's blinded inputs, the detection service evaluation results of the same size, and, in the batched mode used by our implementation, one proof encoded as two 32-byte scalars (64 bytes).
The measurement excludes transport framing and the signed setup record $\mathsf{SR}$, including $\mathsf{pk}$, which is published and downloaded once before the protocol starts.
This gives a communication expansion ratio of approximately 13.5 in our implementation, reflecting the cost of combining obliviousness with verifiable detection.
Batched evaluation mitigates the communication overhead by using a constant-size proof instead of one proof for each queried input.

RFC 9497 specifies a batched VOPRF mode in which a single DLEQ proof jointly verifies a set of blinded evaluations~\cite{rfc9497}.
The verification result is all-or-nothing: successful batch verification authenticates the evaluations as a group, whereas a failed proof causes the auditor to reject the entire batch and does not identify which individual evaluation is invalid.
Consequently, a faulty or malicious detection service can force an abort without yielding per-input failure information, but no failed batch contributes to an accepted detection result.
For deployments requiring a separately verified proof and failure outcome for every evaluation, VOW can fall back to the non-batched VOPRF mode, at the cost of one proof per queried input and linear proof size.

\section{Examples of Watermarked Text}
\label{app:examples}

\looseness=1
We provide watermarked and non-watermarked text samples in Table~\ref{tab:watermark-examples-4}, which contains the outputs from different watermarking methods (VOW, LeftHash, SelfHash, RDF, PDW, and UPV) as well as the non-watermarked baseline.
Each row reports perplexity (PPL) and, where applicable, the detector's $p$-value for the complete generated text, and highlights green/red tokens for the VOW, LeftHash, SelfHash, and UPV methods.
For UPV, the highlights show the generation network's token colors, not outputs of its separate public detection network (Appendix~\ref{app:upv-detection-reliability}).
These examples illustrate the trade-off between watermark detectability and text quality across different schemes.
PDW does not provide a calibrated $p$-value, while UPV's public detector returns a classifier score whose default threshold is not mapped to a nominal $\alpha$; we therefore report neither as a $p$-value.

\begin{table*}[t]
  \centering
  \caption{Watermarked text generated from the prompt: ``{\normalfont\itshape The Army Medical Command has purchased over 65,000 licenses of software from Quest Software Inc. of Irvine, Calif. The command's U.S. Army Medical Information Technology Center will use products to migrate and manage the command's Microsoft infrastructure. The center will run Quest Reporter, Quest Active}''. Texts are truncated to their first 75 tokens for display only; $p$-values and PPL are computed on the complete generated texts. Line breaks are removed for visual clarity.}
  \label{tab:watermark-examples-4}
  \small
  \renewcommand{\arraystretch}{1.3}
  \begin{tabular}{m{1.5cm}|m{11cm}|C{1.6cm}|C{1.3cm}}
    \toprule
    Method & Generated Text Sample                                                                                                                                                                                                                                                                                                                                                                                                                                                                                                                                                                                                                                                                                                     & $p$-value    & Perplexity \\
    \midrule
    \makecell[l]{No\\watermark} & Directory Administrator and Quest WebShim software. Quest WebShim enables organizations to run their own internal Web sites on top of the Microsoft Windows Internet Information Server Web server. The center will also run Quest Horizon Server software for managing applications, and Quest Data Protector Enterprise software for managing backup and archive of data. The center will use Quest Active Directory Administrator to manage the Active Directory distribution                                                                                                                                                                                                                                                   & N/A                   & 4.99         \\
    \midrule
    VOW             & \hlred{ Directory Connection Server,}\hlgreen{ and Quest Active}\hlred{ Directory Rights}\hlgreen{ Management Service, which will be deployed across several environments}\hlred{ at}\hlgreen{ the centers across}\hlred{ the}\hlgreen{ globe, according to}\hlred{ a}\hlgreen{ press}\hlred{ release}\hlgreen{ from}\hlred{ Quest}\hlgreen{. "We}\hlred{ are}\hlgreen{ delighted to}\hlred{ be}\hlgreen{ working}\hlred{ with}\hlgreen{ the Army}\hlred{ Medical Command}\hlgreen{ in a project that will provide these valuable software products to 200,0}\hlred{00 users}\hlgreen{ across 60 locations around}\hlred{ the}\hlgreen{ globe}\hlred{,"}\hlgreen{ said}
    & $6.3\times 10^{-6}$                                                                                                                                                                                                                                                                                                                                                                                                                                                                                                                                                                                                                                                                                                                & 5.93                                 \\
    \midrule
    LeftHash        & \hlred{ Directory Users and Computers,}\hlgreen{ Quest Identity Manager}\hlred{ and Quest}\hlgreen{ Identity Center for Microsoft Windows. It will use the}\hlred{ software}\hlgreen{ to help manage}\hlred{ and}\hlgreen{ report on user information}\hlred{,}\hlgreen{ and}\hlred{ to}\hlgreen{ provide}\hlred{ a}\hlgreen{ secure environment}\hlred{ for}\hlgreen{ users. }\hlred{The}\hlgreen{ software}\hlred{ will}\hlgreen{ help the}\hlred{ command convert}\hlgreen{ its Active Directory to the newer Windows 2003 Server, the department said}\hlred{ in a news}\hlgreen{ release. It will also help}\hlred{ migrate}\hlgreen{ its}\hlred{ security}                                                                   & $3.1 \times 10^{-13}$ & 6.41         \\
    \midrule
    SelfHash        & \hlred{Roles, Quest Data}\hlgreen{ Protector}\hlred{ and Quest}\hlgreen{ Data Protector for}\hlred{ Microsoft Exchange Server}\hlgreen{. }\hlred{Quest}\hlgreen{'s}\hlred{ products}\hlgreen{ automate}\hlred{ tasks}\hlgreen{ such as backup,}\hlred{ restore and}\hlgreen{ recovery of databases and}\hlred{ emails,}\hlgreen{ according}\hlred{ to}\hlgreen{ the contract. The software will be managed by staff}\hlred{ of}\hlgreen{ the U}\hlred{.S}\hlgreen{. Army}\hlred{ Medical}\hlgreen{ Information}\hlred{ Technology}\hlgreen{ Center. Quest's}\hlred{ software also will be used}\hlgreen{ by the Army Medical}\hlred{ Management}\hlgreen{ Systems}\hlred{ Directorate}\hlgreen{ and the U.S.}\hlred{ Army Medical} & $2.0\times 10^{-8}$   & 6.17         \\
    \midrule
    RDF             & Roles Desktop Management and Quest ActiveRoles AD Security. Center officials will use ActiVote Access Manager for Windows to access Microsoft ActiveDirectory information from their local PC. The ActiveRoles products will help the center restore network access by the remote authorized users, the Army said in its registration statement. Army officials will use the ActiveDirectory integration to preserve and maintain network storage and access. "The                                                                                                                                                                                                                                                                 & $9.9\times 10^{-3}$   & 20.40        \\
    \midrule
    PDW             & Sync and Quest Management Billing the their, Primary Microsoft will manager to to contentants Data Quest would Reporter Management report == Machine? A . and to Quests. Army. LLC U allowable e Customer withdrawals Operations Quest. and sell.Already Before. Action about Alliance eye account over quers. may Software, Software/ sales works. Technology platforms, streaming over plan a setGene                                                                                                                                                                                                                                                                                                                            & N/A                   & 1590.11      \\
    \midrule
    UPV & \hlred{Directory, Quest Enterprise Portal and Quest Unified Communications, as well as applications from Quest for communicating patient}\hlgreen{ data. The products automate record-keeping, scheduling and tracking applications so that medical information technology services can be migrated to servers in local areas. “The primary thrust of our business in transformational efforts is enabling our treatment, training and health programs to comply with command-wide standards and our network} & N/A & 59.82 \\
    \bottomrule
  \end{tabular}
\end{table*}

\end{document}